\documentclass[review,times]{elsarticle}
\usepackage{graphicx,colordvi}
\usepackage{bm}
\usepackage{subfig}
\usepackage{xcolor}
\usepackage{listings}
\usepackage{url}

\providecommand{\tabularnewline}{\\}

\begin{document}
\begin{frontmatter}

\title{LFRic: Meeting the challenges of scalability and performance portability in Weather and Climate models}

\author[met]{S.~V.~Adams}
\author[hartree]{R.~W.~Ford}
\author[met]{M.~Hambley}
\author[met]{J.~M.~Hobson}
\author[met]{I.~Kav\v ci\v c}
\author[met,read]{C.~M.~Maynard}
\ead{c.m.maynard@reading.ac.uk}
\author[met]{T.~Melvin}
\author[bath]{E.~H.~M\"{u}ller}
\author[met]{S.~Mullerworth}
\author[hartree]{A.~R.~Porter}
\author[monash]{M.~Rezny}
\author[met]{B.~J.~Shipway}
\author[met]{R.~Wong}

\address[met]{Met Office, UK}
\address[hartree]{STFC Hartree Centre, Daresbury Laboratory, UK}
\address[read]{Department of Computer Science, University of Reading, UK}
\address[bath]{Department of Mathematics, University of Bath, UK}
\address[monash]{Monash University, Melbourne, Australia}

\begin{abstract}

  This paper describes LFRic:\ the new weather and climate modelling
  system being developed by the UK Met Office to replace the existing
  Unified Model in preparation for exascale computing in the 2020s.
  LFRic uses the GungHo dynamical core and runs on a semi-structured 
  cubed-sphere mesh. The design of the supporting infrastructure follows
  object-oriented principles to facilitate modularity and the use of
  external libraries where possible. In particular, a `separation of concerns'
  between the science code and parallel code is imposed to promote
  performance portability. An application called PSyclone, developed at the
  STFC Hartree centre, can generate the parallel code enabling deployment of
  a single source science code onto different machine architectures.
  This paper provides an overview of the scientific requirement, the design
  of the software infrastructure, and examples of PSyclone usage. Preliminary
  performance results show strong scaling and an indication that hybrid
  MPI/OpenMP performs better than pure MPI.  

\end{abstract}

\begin{keyword}
Separation of concerns; Domain specific language; Exascale; Numerical weather prediction
\end{keyword}

\end{frontmatter}

\newpage
\section{\label{sec:intro}Introduction}

The Met Office develops and maintains a large suite of numerical
models that underpins its operational weather and climate research
work. The Unified Model (UM) of the atmosphere is at the heart of this
suite. The UM was developed in the late 1980s and introduced into
operational use in 1990~\cite{gmd-10-1487-2017}. Since its inception
there has been a continuous drive to increase its complexity, accuracy
and compute performance so as to deliver the Met Office's purpose
which is to work at the forefront of weather and climate science for
protection, prosperity and well-being.

In continuing to improve the Met Office's atmosphere model capability
into the future, three challenges were anticipated.

Firstly, for many years much of the improvement in compute performance
of the UM could be obtained by buying high performance computers with
faster processors. However, clock speeds of new generations of standard
CPUs are now often slower than the old. Scaling to more CPUs is
limited by interprocessor communications, and the costs of powering
machines with higher core counts is becoming unsustainable. All these
factors challenge the ability to deliver higher resolution and more
complex numerical models~\cite{gmd-2017-186}.

Secondly, a particular challenge for many atmosphere models, including
the UM, relates to their use of the latitude-longitude (lat-lon) grid.
The convergence of longitude lines at the poles causes numerical
issues that make it hard to scale the model to higher core counts.

Thirdly, looking forward to the future, it is anticipated that HPC
architectures will change radically, and that the diversity of
architectures will increase. Even when porting from one CPU-based
machine to another, there is a cost due to the need to re-tune the
performance for the particular characteristics of the new
machine. With the emergence of more diverse machines such as GPU-based
machines, the cost will only get worse.

In collaboration with academic partners, the Met Office reviewed the
options available to resolve these three issues~\cite{GHP1_CSR}. The
decision was to develop a new dynamical core called
GungHo~\cite{melvin2018} written to run on an unstructured mesh that
avoids the polar singularity problem. A new software infrastructure
would be required as the UM infrastructure will not support an
unstructured mesh. The architecture of the software infrastructure
would impose a separation between scientific code and parallel systems
code -- code that supports parallelism on HPC machines -- so as to
reduce the cost and complexity of porting to new architectures. This
paper describes the novel features of this new model infrastructure
and experiences in developing scientific code within it, and provides
some preliminary compute performance results.

\subsection{The LFRic Roadmap}

LFRic is the name given to the new atmospheric model and to the new
software infrastructure which is being developed to host the GungHo
dynamical core. LFRic is named after the pioneering weather forecaster
Lewis Fry Richardson (~\cite{Lynch2006},~\cite{ForecastFactory}). The
LFRic Roadmap comprises three phases: developing the first version of
the software infrastructure to support the GungHo dynamical core;
extension of the software infrastructure to support implementation of
a full atmosphere model including physics codes; deployment of the
model in operational trials in the lead up to its replacement of the
Unified Model.

Currently LFRic is mid-way through Phase 2, which will end in
2020. The start of Phase 3 marks the point when the LFRic atmosphere
model replaces the Unified model as the Met Office's main target for
science and computational performance improvements. Phase 3 is
expected to end with the deployment of the LFRic atmosphere model in the Met
Office NWP Operational Suite towards the mid-2020s, following which
the first climate configuration will be developed that uses LFRic.

At the core of the LFRic design, the software architecture of the
natural science code imposes a {\em separation of concerns} between
science code and code relating to parallelisation of the model. The
architecture is called PSyKAl after the three layers it comprises:
Parallel Systems or PSy layer code, Kernel code and Algorithm
code. The architecture aims to separate scientific code in the
algorithms and kernels from parallel code within the PSy
layer. Metadata embedded in the scientific code is read by an
application called PSyclone which automatically generates the PSy
layer code. Initially, PSyclone has successfully converted serial code
into OpenMP and MPI parallel code without changing any of the science
code. The design is described in more detail in Section~\ref{sec:SoC}.

The PSy layer code generated by PSyclone includes calls to the LFRic
software infrastructure. The LFRic infrastructure implements a data
model that supports finite element, finite volume and finite
difference model fields, domain decomposition of these fields on
distributed memory platforms, and halo swaps of fields to support
communication of information between distributed memory domains. This
infrastructure is implemented in Fortran and uses Fortran 2003
constructs to impose the separation of concerns.

It is important to be clear that most of the focus of the project so
far has been to support the science requirements of the GungHo
dynamical core and atmosphere model, and to demonstrate the principle
of the separation of concerns within the PSyKAl approach. Performance
results provided in this paper are based on only parts of the
scientific model (the dynamics and individual kernels) and are
therefore preliminary. Currently development of the GungHo dynamics is
still ongoing: important optimisations to its algorithmic performance
such as provision of a multigrid preconditioner are not
complete. Furthermore, while the LFRic infrastructure is capable of
running physics codes copied from the UM, the codes are currently
being pulled in as is, with no consideration of compute performance as
yet.

Additionally, while targeting of future platforms is planned, even now
the ability to test LFRic on some such machines is limited by lack of
sufficient Fortran 2003 support by some compilers for the
object-oriented features being exploited in the design of the LFRic
infrastructure. While lack of compiler support has impacted the
development and testing of LFRic, in the long run the OO design will
better enable the infrastructure to develop and adapt without
impacting scientific code.

The rest of the paper is organised as follows: The GungHo dynamical
core and computational aspects are presented in
Section~\ref{sec:GH}. The software design for the separation of
concerns and PSyKAl API are described in Section~\ref{sec:SoC}. The
model infrastructure and use of libraries is discussed in
Section~\ref{sec:lib}. PSyclone, the code generator is presented in
Section~\ref{sec:psyclone} and the package developed to apply linear
solvers and preconditioners is presented in Section~\ref{sec:Solver}.
Finally a scaling analysis is presented in Section~\ref{sec:scal} and
conclusions drawn in Section~\ref{sec:con}.

\section{\label{sec:GH}GungHo}

The dynamical core in an atmospheric model is responsible for
simulating those fluid dynamical processes that are resolved by the
underlying mesh. It is then coupled to a suite of subgrid physical
parametrization schemes for processes that are not resolved (such as
cloud microphysics and convection) and those that act as subgrid
diabatic sources (such as longwave and shortwave radiative
heating/cooling). For the purposes of this paper only the dry
dynamical core without the subgrid processes is considered.

\begin{figure}[ht]
\centering\includegraphics[width=0.6\linewidth]{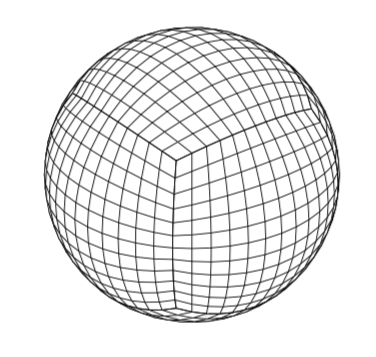}
\caption{\label{fig:cubed-sphere} Equi-angular cubed-sphere mesh as
used in GungHo with $12x12$ subdivisions per face, 
referred to as a $C12$ mesh. This gives $864$ columns of cells.}
\end{figure}

The GungHo dynamical core has been developed to replace the current ENDGame
dynamical core in the UM. The GungHo model seeks to replicate the
accuracy and stability of ENDGame whilst replacing the regular lat-lon
grid with a quasi-uniform grid. This avoids the problems associated
with the polar singularities in ENDGame resulting from the
convergence of meridians at the poles. The mesh used to develop GungHo
within LFRic is an equi-angular cubed-sphere, shown in Figure~\ref{fig:cubed-sphere}
~\cite{NairEtal2005}. This has the benefit of near uniform 
resolution across the globe and a maximum/minimum edge length of
$\approx 1.3$ achieved by using only quadrilateral cells.  However,
this quasi-uniform mesh comes at the expense of losing orthogonality
(the line between two neighbouring cells centres is not, in general,
perpendicular to the edge shared between the two cells). Additionally,
the two polar singularities of the lat-lon grid have been replaced by
the eight corners of the cubed sphere, where only three cells meet at a
vertex instead of the usual four. The lack of orthogonality and the
presence of the corner singularities need to be taken into account when
choosing an appropriate numerical method to avoid errors being dominated
by the corners.

To generate the 3D mesh this 2D cubed-sphere mesh is extruded in the radial 
direction to form a spherical shell of cells. Where orography is present a terrain
following radial coordinate is used such that every column contains the
same number of cells. The horizontal mesh is treated as unstructured,
such that it could easily be changed, for example, to an icosahedral mesh as
considered in~\cite{staniforth2012}.
The vertical mesh is structured and directly addressed, so that
a data point is addressed as \textit{map(df,col)+k}, where \textit{col}
is the horizontal index of the column and \textit{k=0,...,nlayers-1} is
the index of the vertical layer; \textit{df} is the index of a particular
data point within the cell, allowing multiple data points within a single cell, and 
finally \textit{map} is an array containing the address of data points at 
the bottom of the model column.
Using an extruded mesh of this type with direct access and data contiguity
in the vertical, the cost of the indirectly addressed horizontal index can 
be offset (see~\cite{MacDonaldEtal2011}) provided enough layers are used. 
Since the current UM uses~$O(100)$ layers the cost of the indirection is minimal.

\subsection{Formulation\label{sec:sub:formulation}}

The GungHo dynamical core solves the Euler equations for a perfect gas in a 
rotating frame
\begin{eqnarray}
\frac{\partial\mathbf{u}}{\partial t} & = & -\left(2\bm{\Omega}+\nabla\times\mathbf{u}\right)\times\mathbf{u} - \nabla\left(\frac{1}{2}\mathbf{u}\cdot\mathbf{u} + \Phi\right) - c_p\theta\nabla\Pi,\label{eq:momentum}\\
\frac{\partial\theta}{\partial t} & = & - \mathbf{u}\cdot\nabla\theta,\label{eq:energy}\\
\frac{\partial\rho}{\partial t} & = & - \nabla\cdot\left(\mathbf{u}\rho\right)\label{eq:continuity},
\end{eqnarray}
the system is closed by the equation of state
\begin{equation}
\Pi^{\frac{1-\kappa}{\kappa}} = \frac{R}{p_0}\rho\theta.\label{eq:eos}
\end{equation}
These constitute a set of coupled non-linear Partial Differential
Equations (PDEs) for the vector wind $\mathbf{u}$, 
the density $\rho$, potential temperature $\theta$ and Exner pressure $\Pi$. 
Additionally: $\bm{\Omega}$ is the rotation rate; $\Phi$ is the geopotential; 
$p_0$ is a reference pressure; $R$ is the gas constant; $c_p$ is the specific 
heat at constant pressure and $\kappa\equiv R/c_p$.

\subsection{Spatial Discretisation\label{sec:sub:spatial}}

\begin{figure}
\vspace{-2cm}
\centering
\vspace{-1cm}
\subfloat[]{\includegraphics[width=0.5\linewidth]{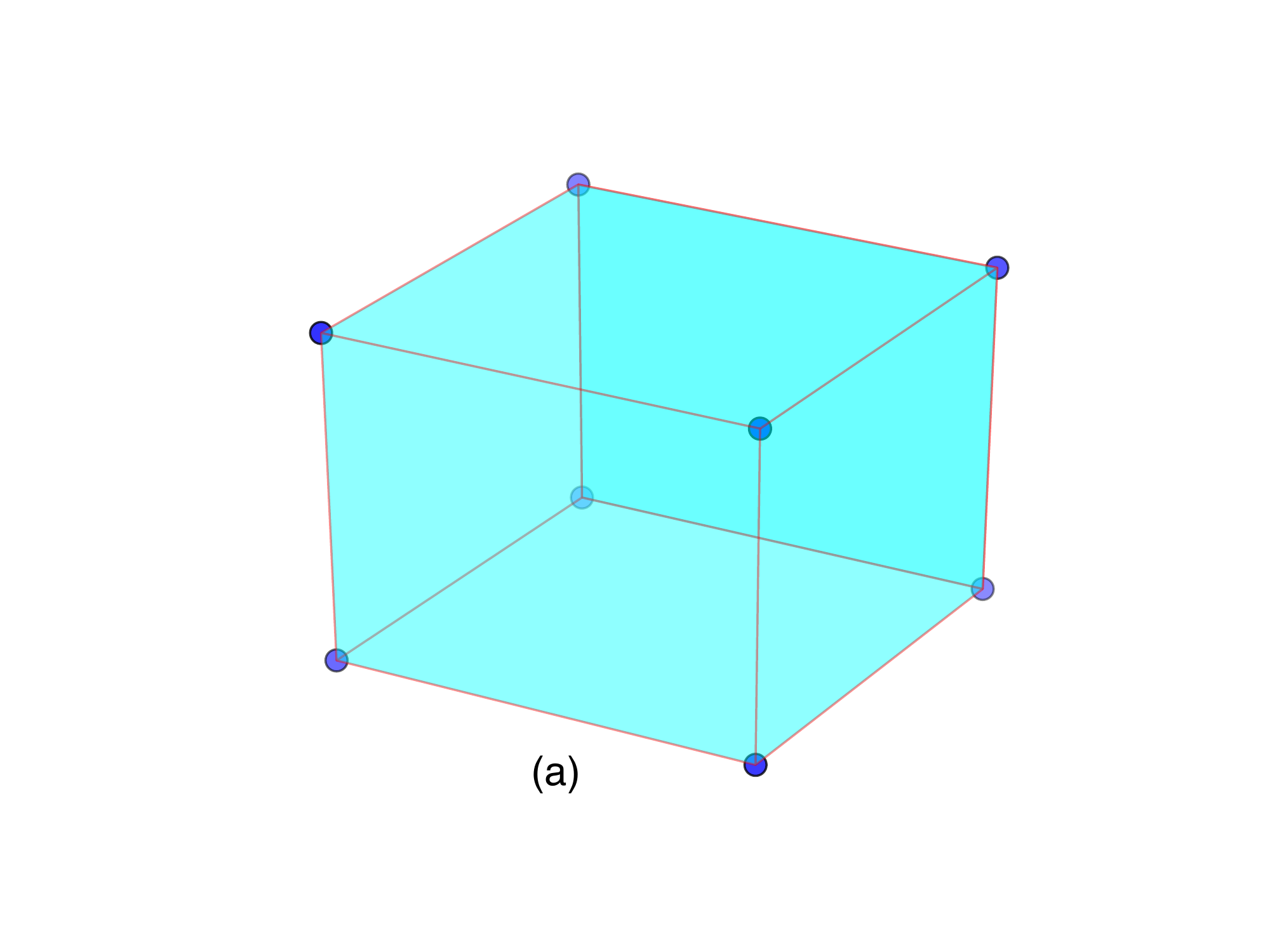}}%
\subfloat[]{\includegraphics[width=0.5\linewidth]{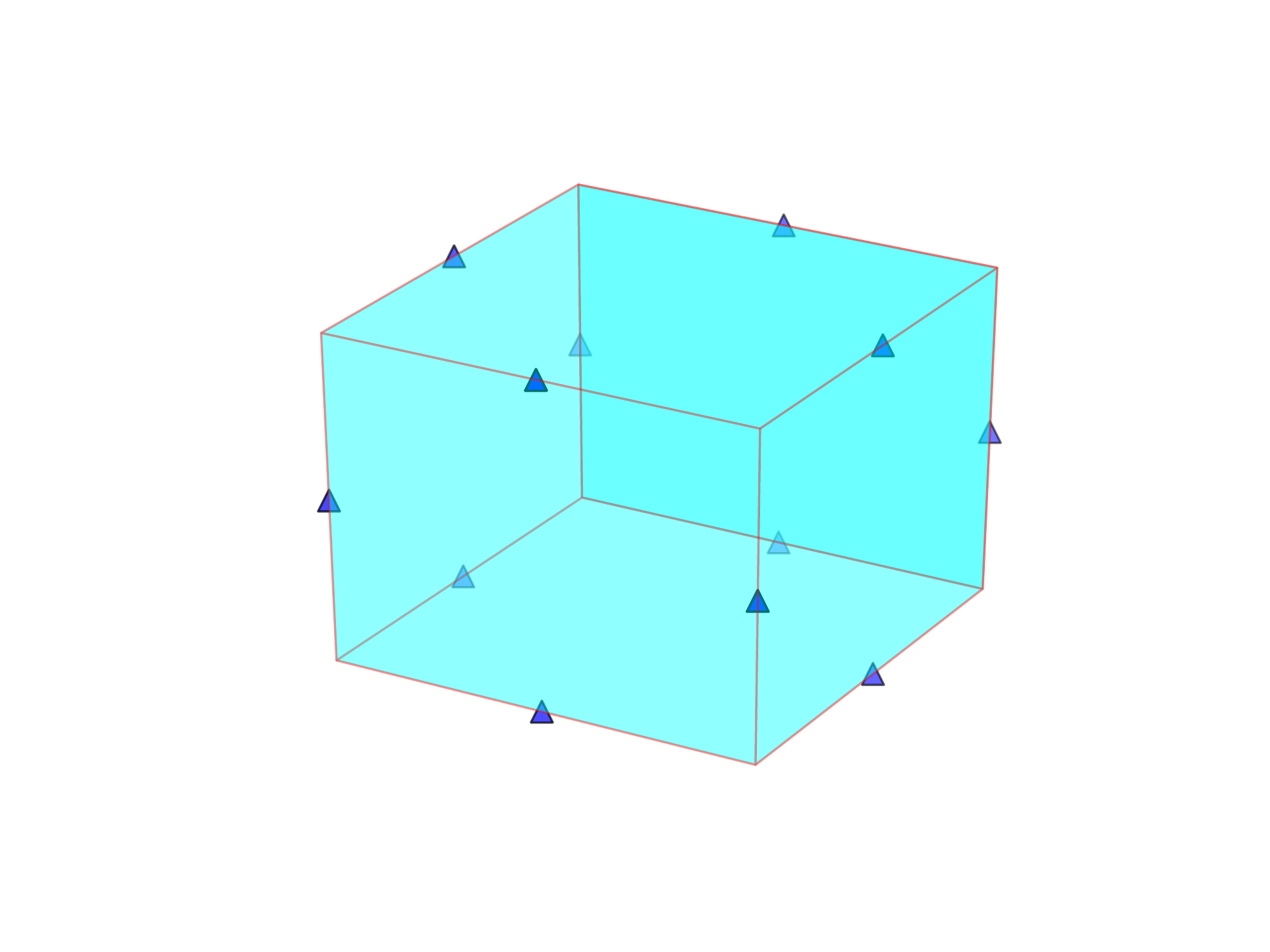}}\\
\subfloat[]{\includegraphics[width=0.5\linewidth]{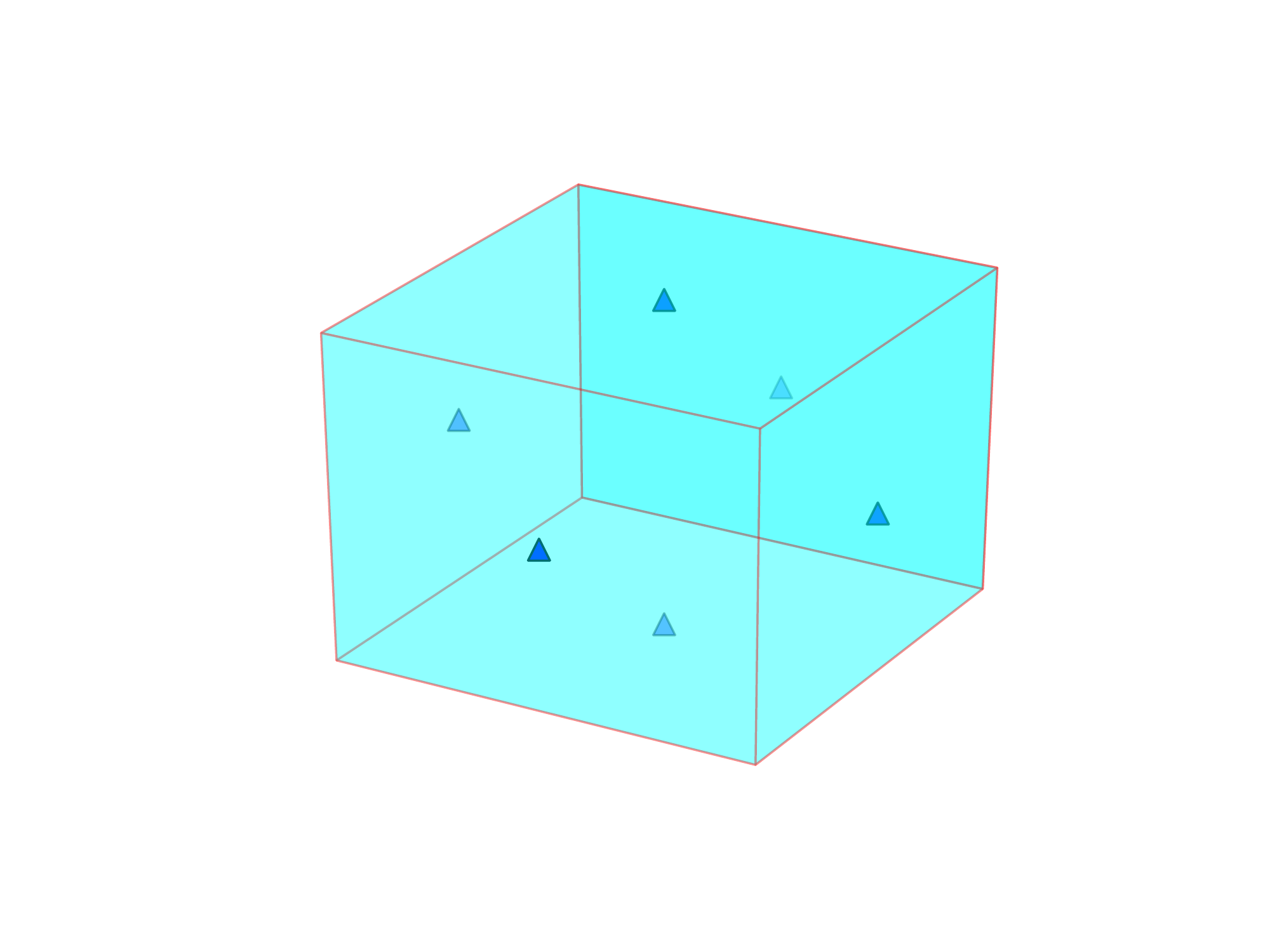}}%
\subfloat[]{\includegraphics[width=0.5\linewidth]{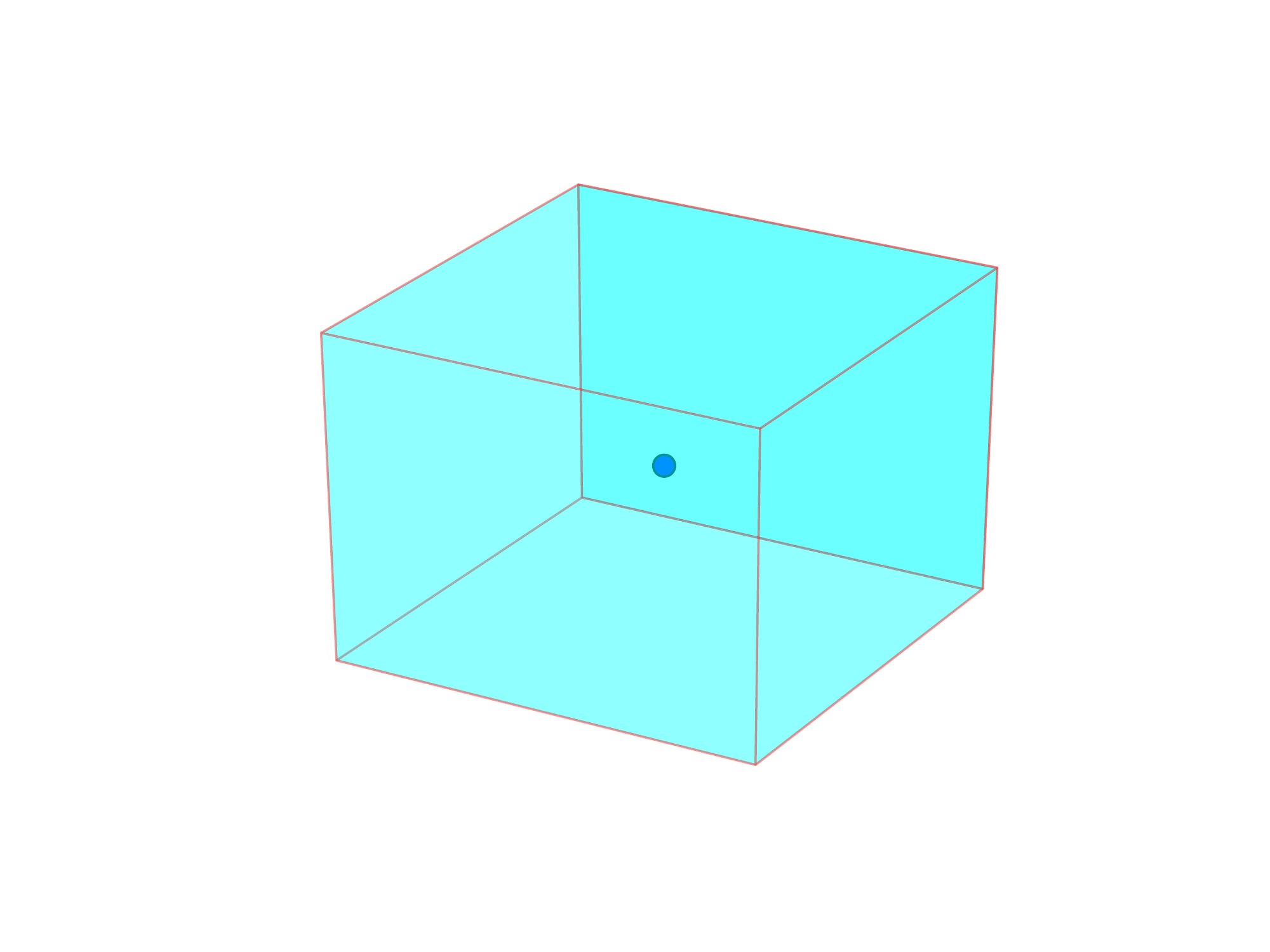}}%
\caption{\label{fig:fem-spaces} Location of degrees of freedom at
  $p=0$, (a) $Q_{1}$, (b) $N_0$, (c) $RT_0$ and (d) $Q_0^D$ finite element spaces.
Circles correspond to scalar degrees of freedom and triangles to vector degrees
of freedom.}
\end{figure}

In order to maintain a similar
accuracy to ENDGame on a quasi-uniform mesh a mixed Finite Element Method is 
used~\citep{cotter2012, natale2016}. This gives the finite element equivalent 
of the C-grid-Charney-Phillips staggering~\citep{charney1953numerical, arakawa1977computational}
used in ENDGame, but without relying
on the orthogonality of the mesh for numerical consistency. The mixed FEM
is very general in terms of the order of approximation and shape of the 
underlying mesh, allowing the method to be tailored to specific needs of the 
application. The mixed FEM involves defining a number of finite 
element spaces and differential mappings between them. The particular family of 
finite element spaces~\citep{boffi2013}, for a given polynomial order $p$, used 
in GungHo are: $Q_{p+1}$ containing continuous 
point wise scalar quantities; $N_p$ containing circulation vectors that have
continuous tangential components; $RT_p$ containing flux vectors that have 
continuous normal components; $Q_p^D$ containing discontinuous volume 
integrated scalars; along with a horizontally discontinuous vertically continuous 
space~\citep{natale2016} for $\theta$ to mimic the Charney-Phillips grid staggering 
used in ENDGame. Here continuous means that neighbouring cells share degrees of freedom 
(hereafter dofs) located on the shared entities ({\em e.g.} faces, edges, vertices).
In practice the lowest order spaces $p=0$ are used, the resulting location of the dofs
for these spaces are shown in Figure~\ref{fig:fem-spaces}. 

The velocity field $\mathbf{u}$ is placed in the $RT_0$ space; the vorticity 
$\xi\equiv\nabla\times\mathbf{u}$ is placed in the $N_0$ space; density, $\rho$, 
and Exner pressure, $\Pi$, are placed in the $Q_0^D$ space; as mentioned above 
the potential temperature, $\theta$, is placed in a scalar-space corresponding 
to the vertical components of $RT_0$ with dofs located in the
centre of the top and bottom faces of a cell. See~\cite{melvin2018} for a full
description of the discretisation used in the GungHo model.

\subsection{Advection\label{sec:sub:advection}}

In order to achieve accuracy similar to that of ENDGame it is important to have 
a high-order approximation to the scalar advective terms: $\mathbf{u}.\nabla\rho$ 
and $\mathbf{u}.\nabla\theta$. In ENDGame these are evaluated using a 
semi-Lagrangian formulation that involves computation of the trajectories that 
the fluid has taken over a timestep, followed by a high-order multi-dimensional 
interpolation of the scalar field to the origin of the trajectory. 
The non-local nature of the trajectory computation can 
lead to significant communication costs if the trajectory crosses over processor 
boundaries, which often happens near the poles of the lat-lon grid.

In GungHo this semi-Lagrangian scheme is replaced by an Eulerian finite-volume 
method of lines advection scheme that maintains inherent local conservation of mass.
The method fits a high-order upwind polynomial over a number of
neighbouring cells and evaluates this at a fixed point to compute the advective 
term. This only requires a local computation and the stencil is fixed, reducing the 
amount of computation needed. However, in order to obtain a stable scheme it is 
wrapped in a multi-stage evaluation, meaning that the advection update needs to be 
computed a number of times per iteration of the iterative-semi-implicit scheme
(see Section~\ref{sec:sub:timestepping}); nominally 3 stages are used. 
Future work will evaluate the use of flux-form Semi-Lagrangian scheme (COSMIC)~\cite{Leonard1996} 
to replace the current scheme. This carries many of the 
benefits of the semi-Lagrangian scheme used in ENDGame but without the 
communication costs (due to use of a quasi-uniform mesh) as well a simplification 
to the computation from using a dimensionally split formulation that only requires 
one-dimensional interpolation.

\subsection{Time-stepping\label{sec:sub:timestepping}}

As in ENDGame, to facilitate the use of long time-steps, a two time-level 
iterative semi-implicit time-stepping scheme is used. This requires, within each 
time-step, a non-linear Picard iteration to update the non-linear and advection terms 
and at each iteration a large sparse linear system is solved, which is formed by 
use of a quasi-Newton method,
\begin{equation}
\mathcal{L}\left(\mathbf{x}^n\right)\mathbf{x}' = \mathcal{R}\left(\mathbf{x}^{(k)}\right),\label{eq:quasi-newton}
\end{equation}
for the increment $\mathbf{x}'\equiv\mathbf{x}^{(k+1)}-\mathbf{x}^{(k)}$ on the $k^{\rm{th}}$ 
estimate from the Picard iteration of the prognostic variables $\mathbf{x}\equiv\left(\mathbf{u},\,\theta,\,\rho,\,\Pi\right)$. 
The linear system $\mathcal{L}\left(\mathbf{x}^n\right)$ is chosen as to contain the terms necessary for stability of 
the fast acoustic and gravity waves as in~\cite{QJ:QJ2235} and is formed from a linearisation about the previous 
timestep fields $\mathbf{x}^n$. The method used to solve this linear system is detailed in Section~\ref{sec:Solver}. 
An overview of the time-step is given in Table~\ref{tab:timestep}.
\begin{table}
\begin{centering}
\begin{tabular}{l}
\hline 
\textbf{do} $n=1,N$ (begin time-step loop)\tabularnewline
\hspace{0.5cm}Set $\mathbf{x}^{(1)} = \mathbf{x}^n$\tabularnewline
\hspace{0.5cm}Compute $\mathcal{L}\left(\mathbf{x}^n\right)$\tabularnewline
\hspace{0.5cm}Compute $R\left(\mathbf{x}^n\right)$\tabularnewline
\hspace{0.5cm}\textbf{do} $k=1,K$ (begin Picard loop)\tabularnewline
\hspace{1.0cm}Compute advective terms $R^A\left(\mathbf{x}^{(k)},\mathbf{x}^{(n)}\right)$\tabularnewline
\hspace{1.0cm}Compute $R\left(\mathbf{x}^{(k)}\right)$\tabularnewline
\hspace{1.0cm}Set $\mathcal{R}^{(k)} = R^n + R^{(k)} + R^A$\tabularnewline

\hspace{1.0cm}Solve $\mathcal{L}\left(\mathbf{x}^n\right)\mathbf{x}' = \mathcal{R}^{(k)}$\tabularnewline
\hspace{0.5cm}\textbf{end do}\tabularnewline
\hspace{0.5cm}Set $\mathbf{x}^{n+1} = \mathbf{x}^{(K)}$\tabularnewline
\textbf{end do}\tabularnewline
\hline
\end{tabular}
\end{centering}
\caption{\label{tab:timestep}Overview of a single timestep in the GungHo dynamical core, typically $K=4$.}
\end{table}

\section{\label{sec:SoC}Separation of Concerns}

Science applications in general and weather and climate codes in
particular are written in high-level languages such as Fortran or
C/C++. Fortran is commonly employed for weather and climate
codes as it is especially suited to numeric computation. Using
such a high-level language, an algorithm is written to solve a
mathematical problem without consideration for the processor
architecture. The compiler generates machine-specific instructions
and can, in principle, make optimisation choices to exploit the
architecture of different processors.

This abstraction of a separation of concerns between mathematics code
and machine code is powerful. It enables the portability of
science code to different processor architectures, and allows the
application to exploit a significant fraction of the peak
performance of the processor. Whilst many science applications may
contain code optimisations in performance-critical sections (for
example, blocking or tiling loop nests to better utilise cache memory),
in general these applications have relied on clock
speed increases between processor generations to increase performance.

Increases in processor speed between generations ceased more than a decade
ago due to the ending of Dennard scaling~\cite{dennard}. Instead,
successive generations of processors have had increasing numbers of processor cores per
socket. Science applications typically have already been adapted to run on multiple
nodes to exploit supercomputers with the distributed memory and data parallelism
typically expressed via MPI. Multi-core nodes present
additional opportunities and challenges to applications in terms of exploitation of
shared address space within a node. Heterogeneous compute nodes such as
CPU + GPU with distinct memory spaces require further additional programming
models to enable proper exploitation of the available compute performance.

A great number of programming models exist. For distributed memory there are, 
for instance, MPI and Partitioned Global Address Space (PGAS) languages, such 
as Co Array Fortran, Unified Parallel C, Chapel and GASPI. For threaded and 
shared-memory parallelism there are directive-based solutions such as OpenMP 
and OpenACC as well as languages such as CUDA and OpenCL. However,
the programming models lag behind in development of the rapidly evolving computer
architectures. Particular models lack feature or processor coverage
making the choice of programming model difficult. These issues are
especially difficult for weather and climate applications with their long development
cycles. 

Many applications have adopted an MPI + X model, where X is one or
more of the programming models mentioned in the previous paragraph. This
is problematic for several reasons. The programming models are not
only different, they are different types of model: languages, language
extensions, libraries and directives, some of which are architecture
specific. Worse, the interaction between them is outside the scope of
any model and may, for example, depend on control by the batch scheduling
system. The applications may require a different X for
different architectures and even different data layouts and
loop nest order. All these parallel and performance features
break the data abstraction of the separation of concerns between the
maths/science code and architecture-specific code.

\begin{figure}
\centering\includegraphics[width=0.8\linewidth]{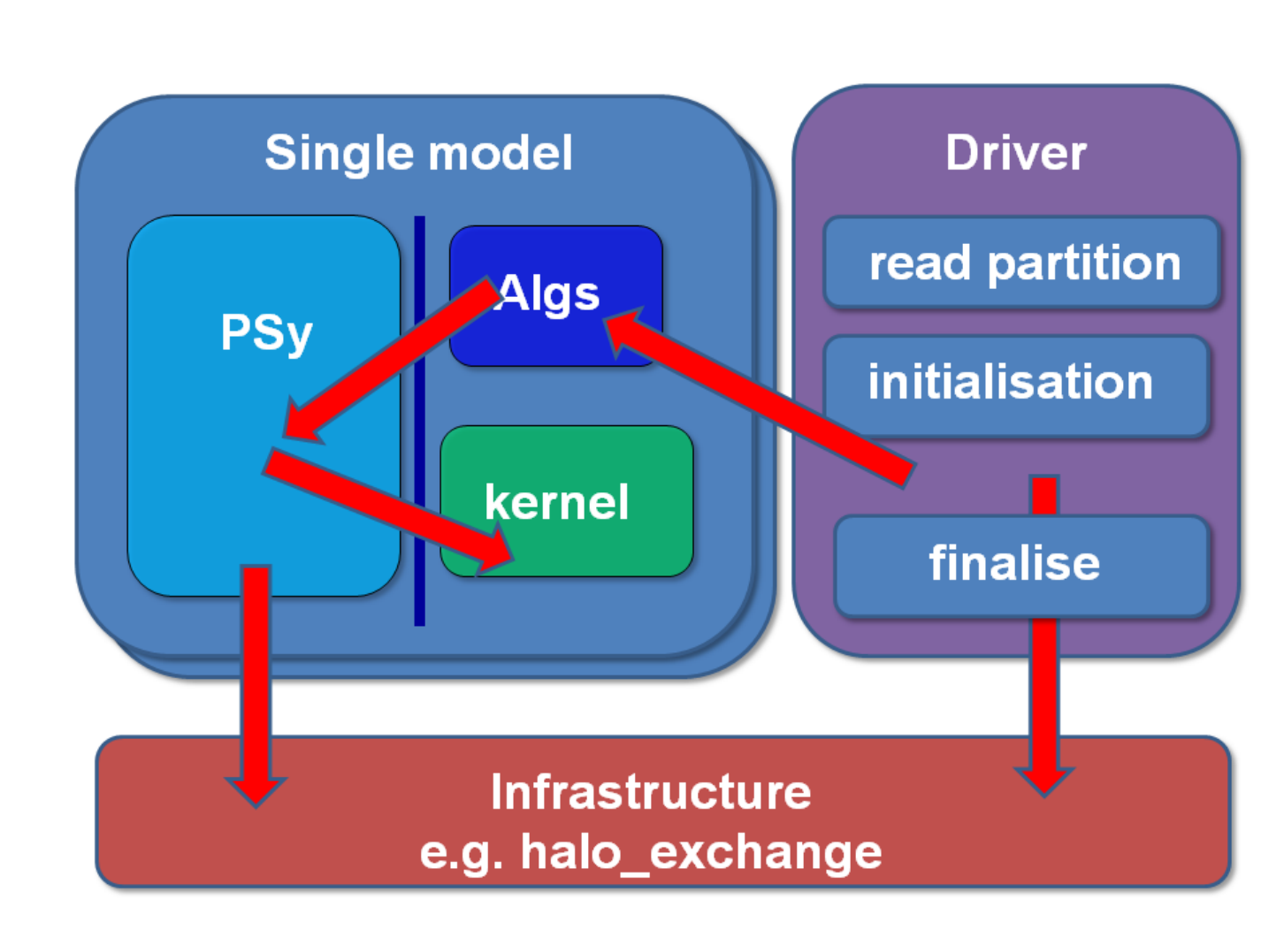}
\caption{\label{fig:psykal} Schematic diagram of the PSyKAl layered
  architecture. A single model could be for example, the atmospheric
  model. The blue vertical line between the PSy layer and the
  Algorithm and Kernel layers represents a separation of the science
  and parallel code.}
\end{figure}

The design for the LFRic model is based upon a computational science
report from Phase 1 of the GungHo project~\cite{GHP1_CSR}. By
employing a layered software architecture, the complex parallel code
can be kept separate from the science code. As outlined in 
Section~\ref{sec:intro}, the software is separated into three layers.
Algorithms are expressed as operations on global, data-parallel field objects,
in the top algorithm layer. The middle PSy layer (Section~\ref{sec:intro}) 
contains the looping over the horizontal field and is where the data parallelism is expressed. 
The bottom layer comprises the kernels which encode the operation invoked in the
algorithm layer. Kernels are written to operate on a single vertical column of cells. 
Shown in Figure~\ref{fig:psykal} is a schematic diagram of this
layered software architecture. The red arrows indicate the control flow
through the different layers.

The APIs between the layers are tightly controlled. The PSy layer can
be called only from the algorithm layer by an {\em invoke} procedure
call. The permitted arguments to these invoke procedures are a restricted
set of LFRic objects, such as fields, that the kernels will operate on. The
kernels can be called only from the PSy layer. The PSy layer code
unpacks information from the LFRic objects including data
arrays, scalars for loop bounds, and supporting arrays of both reals
and integers for the Finite Element Method.
The API is implemented using a Fortran 2003 object
orientation coding style. By restricting the
domain of the problem from all numerical mathematics to only that
required (in the first instance) in the GungHo dynamical core it is
possible to constrain the parallelism to the PSy layer.

Compilers in general have to be conservative about the assumptions
they can make in order to guarantee correctness. However, a further
advantage of the DSL approach is that developers can
explicitly express the domain knowledge which is not possible in a
standard, high level language. In the case of the LFRic model, data
access descriptors are employed to say whether access to a field is
{\em read}, {\em write}, {\em readwrite} or {\em increment}. The latter 
two discriminate between fields that do not share degrees 
of freedom between columns ({\em e.g.} density defined on 
$Q_p^D$ function space, see Section~\ref{sec:sub:spatial} for more details) 
and the ones that do ({\em e.g.} velocity defined on $RT_0$ space, 
Section~\ref{sec:sub:spatial}). This information is embedded in
Fortran as part of the kernel. Other information encoded in this way
is the loop iterator (for example, the choice to loop over horizontal {\em cells}), whether the kernel
operates on a field which lives on a specific function space, whether
any of the fields live on common function spaces and for FEM kernels
whether the kernel uses any basis functions. 

The restricted domain, strict control of the APIs between layers and,
critically, the data access descriptors and other kernel metadata
allow a set of rules to be derived to construct the PSy layer. The
code itself can therefore be automatically generated, and a Python
code parser, transformer and code generator called PSyclone has been
written to generate it. More details of PSyclone can be found in
Section~\ref{sec:psyclone}. The generated PSy layer code is Fortran
and contains calls to the LFRic infrastructure (see
Section~\ref{sec:lib} for more details) to access functionality such
as: MPI distributed memory parallelism, dereferencing the Fortran 2003
objects, looping over the horizontal mesh elements and calling the
kernels themselves.  PSyclone can also perform optional
transformations to target other programming models such as OpenMP or
OpenACC.

\subsection{\label{sec:sub:implement}Implementation}

The algorithm layer code is written in Fortran 2003. However, this
code is not actually compiled; it is parsed and processed by PSyclone
first. Most of the Fortran is left as written. The exception to this
is the call to any {\em invoke} procedure. The invoke procedure does not 
exist as such in the source code. Instead, an invoke means {\em execute these
kernels looping over the chosen entities}, hence instructing PSyclone on how to 
generate parallel code. PSyclone can parallelise this horizontal looping with MPI,
OpenMP or both. Shown in the code fragment below is an invoke
procedure call from the algorithm layer.

\begin{lstlisting}[language=Fortran,caption={Code fragment showing an
invoke procedure from the Algorithm layer},label={lst:invoke}]
 call invoke( setval_c(v(m), 0.0_r_def), &
              matrix_vector_kernel_type(v(m), s(m), mm), &
              enforce_bc_kernel_type( v(m) ) )
\end{lstlisting}

There are three kernels in this invoke. The first \verb+setval_c+ is
an example of a {\em point-wise} kernel. The same operation, setting all
the values of the field to the same scalar, is applied to each dof so no 
FEM structure is required. This is an example of a {\em built-in} operation 
generated in-place in the PSy layer by PSyclone. The second
kernel is written as a call to the constructor of the kernel type, so that it is
valid Fortran. The arguments to this call \verb+v+ and \verb+s+ are
{\em fields} and \verb+mm+ is an {\em operator}, in this case a mass
matrix.

PSyclone replaces the code for an invoke to the kernels with a call to a
generated procedure in the PSy layer, with the fields and operators as
arguments. Shown in Listing~\ref{lst:PSy1} is a code fragment from the generated PSy layer for
the call to the matrix-vector kernel in invoke shown in Listing~\ref{lst:invoke}.

\begin{lstlisting}[language=Fortran, numbers=left,caption={Code 
fragment of the generated PSy layer with distributed memory support},label={lst:PSy1}]
 CALL v_proxy%halo_exchange(depth=1)
 IF (s_proxy%is_dirty(depth=1)) THEN
   CALL s_proxy%halo_exchange(depth=1)
 END IF 
 DO cell=1,mesh%get_last_halo_cell(1)
   CALL matrix_vector_code(cell, nlayers, v_proxy%data, &
      s_proxy%data, mm_proxy%ncell_3d, mm_proxy%local_stencil, &
      df_any_space_1_v, undf_any_space_1_v, &
      map_any_space_1_v(:,cell), &
      ndf_any_space_2_s, undf_any_space_2_s, &
      map_any_space_2_s(:,cell))
  END DO 
\end{lstlisting}
Lines 1-4 are the distributed
memory calls to the infrastructure for a halo exchange and
testing/setting flags to indicate whether a halo (and to what depth)
has been updated. Lines 6-12 are the looping over the horizontal mesh
and the procedure call to the kernel itself. The arguments are simple
scalars for sizes and loop counters, the data arrays and (in this example)
the indirection maps.

Shown in the code fragment below is the kernel metadata for the
matrix-vector kernel invoked in the algorithm layer.
\begin{lstlisting}[language=Fortran, numbers=left,caption={Code
fragment showing kernel metadata for the matrix-vector operator kernel},label={lst:metadata}]
type, public, extends(kernel_type) ::  &
                    matrix_vector_kernel_type
  private
  type(arg_type) :: meta_args(3) = (/ &
       arg_type(GH_FIELD,    GH_INC,  ANY_SPACE_1), &
       arg_type(GH_FIELD,    GH_READ, ANY_SPACE_2), &
       arg_type(GH_OPERATOR, GH_READ, ANY_SPACE_1, &
       ANY_SPACE_2)  /)
  integer :: iterates_over = CELLS
contains
  procedure, nopass ::matrix_vector_code
end type
\end{lstlisting}
The metadata is embedded in Fortran so that no special comments or
other mark-up is required. The access descriptors shown above are key
to deriving the appropriate rules for generating the code in the PSy
layer. The \verb+arg_type+ has three components. Firstly, the type of
the data object, in this case fields and operators (an
operator is a mapping between one function space and another). Secondly,
the data access pattern for the data object. For example, \verb+GH_INC+ means the data is
incremented. Thirdly, which function space the data lives on. In this case, the
kernel is general and can operate on any function space, hence, the
\verb+any_space_n+. Whilst the $n$ can be any number, the function space of the
first field must correspond to the first function space of the operator and the
function space of second field must correspond to the second function
space of the operator.

\section{\label{sec:lib}Infrastructure}

As shown in Figure~\ref{fig:psykal}, the PSyKAl layered
architecture is supported by the LFRic infrastructure. The LFRic infrastructure provides
functionality such as distributed memory support (halo exchanges),
colouring for OpenMP threading and the provision of loop bounds that the
generated PSy layer will use.

\begin{figure}
\centering\includegraphics[width=0.8\linewidth]{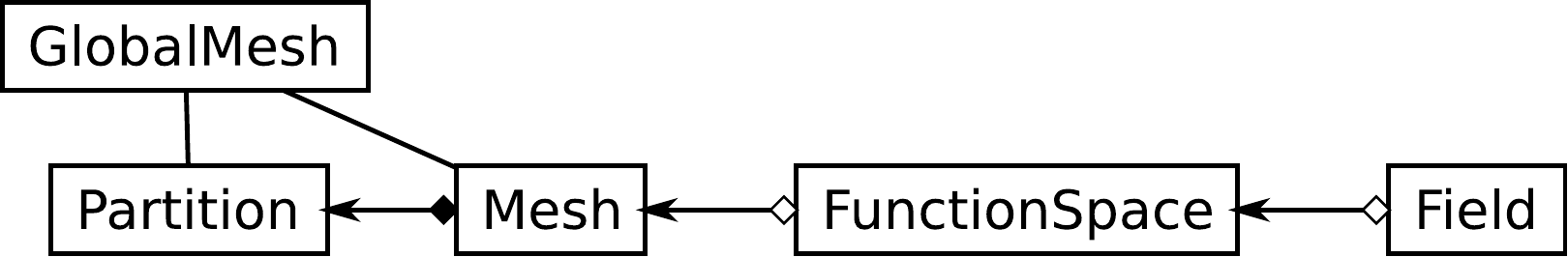}
\caption{\label{fig:objects} A Unified Modelling Language (UML) 
class diagram of the family of objects used
in the LFRic infrastructure to support the field object.}
\end{figure}

Within the development of LFRic there has been a conscious effort to
provide the infrastructure following an object-oriented approach. As
such, much of the infrastructure is provided through a family of
objects that ultimately support the field object that is used within the
science code. The objects are shown in
Figure~\ref{fig:objects}. Two-dimensional topological and positional
information about the global mesh is read into the global model object.
This is passed to a partitioner to generate a partition object that
describes what portion of the global mesh the local processing element
will be working on. The global mesh and the partition objects are then
combined to form a mesh object that extrudes the two-dimensional mesh
information in the vertical for just the part of the mesh that will be
used on the local processing element to form a local, three-dimensional
representation of the mesh. The way the data that is held within the
field object relates to the mesh is described by the function space.

The LFRic infrastructure has been designed to be mesh-agnostic and
thus most of the LFRic infrastructure supports a generic horizontally
unstructured mesh. The partitioner is currently implemented to provide
optimised partitioning specifically for cubed-sphere and planar meshes only.
If a different mesh is required, the only infrastructure change required
to support the new mesh is a relatively straightforward extension of the
partitioner code. 

Most of the infrastructure is accessed through the field object. So, for
example if a halo exchange is required on a field, the halo exchange
method on the field object is called. This infrastructure
API is used by the PSy layer to access information and functionality.

\subsection{\label{sec:sub:distmem}Distributed Memory Optimisation}

Distributed memory parallelism is achieved by partitioning the full
(global) domain into smaller sections (partitioned or local domains).
Information is passed between the tasks through the use of halo regions.
Data in these areas are provided from other partitions that own that
section of the domain.

Performing halo exchanges is not unique to LFRic. Running large parallel
weather and climate model codes has been done for many years, with halo
exchanges being required to communicate updates between the partitions of 
data. However, the following aspects of the data model within the LFRic 
infrastructure make implementing distributed memory parallelism using 
halo exchanges particularly challenging:

\begin{enumerate}

\item LFRic is designed to support horizontally unstructured meshes,
which means it has to support non-uniform shaped partitions with complex
shaped halo regions;

\item Because the domain is horizontally unstructured, moving from one
data point to a horizontally adjacent data point requires using a lookup
table, which is slower than using direct addressing. In order to recover
some code performance, fields are laid out in memory so that vertically
adjacent data points are next to each other, and the looping structure
within LFRic kernels is vertical looping innermost;

\item As noted in Section~\ref{sec:sub:spatial}, the discretisation being used within
LFRic leads to dof data being held on all the different entities within the mesh: 
cell volumes, faces, edges and vertices. A partitioning scheme that partitions
the domain along cell boundaries leads to partitions that must share the
dofs that are located on the boundaries of the domain. Having to assign
ownership of these shared dofs to a particular partition adds to the
complexity of setting up the halo exchanges.

\end{enumerate}

The halo exchanges are achieved by passing information between processes 
using MPI. For performance and ease of use reasons, the MPI library is not 
called directly, instead a layer between the model and MPI is introduced. 
At initialisation time, the layer generates the 
communication routing tables required to perform a halo
exchange. These tables  
are then reused every time a halo exchange is required.

In order to generate the routing table, each rank needs to supply the
following information to the library:

\begin{itemize}

\item A list of all the global dof ids of all the locally owned dofs;

\item A list of all the global dof ids of all the dofs in the halos.

\end{itemize}

Initially, the LFRic project used the infrastructure library component
of the Earth System Modelling Framework (ESMF)~\cite{ESMFDocs} to
provide the layer between the model and MPI, however it was not particularly
well-suited to the higher order finite-element fields used in LFRic . Although
the routing tables it generated were optimal (the halo exchanges called
during timestepping were very fast), only a small part of the ESMF framework
was actually being used. This made it a rather heavyweight dependency.
The design of the model framework to support the `separation of concerns'
led to all the halo exchange functionality being neatly encapsulated in
the PSy layer. Therefore it was relatively easy to try a different library
to provide the intermediate layer. A switch was made to use the YAXT
(Yet Another eXchange Tool)~\cite{YAXTDocs} library. This provides
halo exchanges during time stepping that are as fast as those provided
by ESMF, via a lightweight library.

The API for supporting distributed memory optimisation is very simple. A
halo exchange function is provided on every field and a flag is
maintained by the infrastructure to track whether data held in the halo
cells is up to date with the owner of that data (`clean' halos) or
whether a parallel computation has updated the field on the owning
partition and the local halos are out of date (`dirty' halos). When the
PSy layer determines that a halo exchange may be required, it first
checks the state of the halos and if (and only if) they are `dirty' it 
calls the halo exchange functionality.

When the LFRic project was initiated, a survey of other models and
systems found no other infrastructures or data models that would
support the particular combination of higher order FEM methods and
unstructured meshes. Since then, the ATLAS
library~\cite{DECONINCK2017188} has emerged which in time may
potentially be considered as an alternative data model capable of
delivering LFRic requirements.

\subsection{\label{sec:sub:sharedmem}Shared Memory Optimisation}

As with all the parallelisation within LFRic, the shared memory
parallelism is implemented by inserting directives in the PSy layer. 
The finite-element formulation makes use of data on mesh entities
shared by more than one cell. Two different threads working on
adjacent cells may try writing to a dof shared between the cells at the
same time.
{\em Graph colouring} is applied to the mesh so that no cells of any one
colour share dofs. This means parallel threads can run over the whole
set of cells within a particular colour and safely write dof values
without contention.
The information and functionality to support
colouring within the shared memory parallelism is provided by the LFRic
infrastructure.

When the mesh object is created, it is coloured by the infrastructure
and information about the colouring is stored, so it can be used from
within the PSy layer. The infrastructure provides the number of colours,
the number of cells in each colour and which cells are present in each
colour. Shared memory parallel directives are placed in the PSy layer
and the looping can be made safe using the colouring information
provided by the infrastructure. 

\subsection{\label{sec:sub:io}Parallel I/O}

A consequence of models running over many thousands of cores is a requirement
for scalable parallel I/O. A computationally optimised model is of little practical
operational use if I/O then becomes a limiting factor. The LFRic project has
decided to investigate options for I/O alongside the computational
and infrastructure development for three main reasons. Firstly, the requirement
to be able to run science assessment jobs on large numbers of cores and obtain the
output efficiently. Secondly, a key aim of LFRic is scalability and therefore it is
helpful to be able to monitor the impact of I/O on this as the infrastructure and
science develops. Thirdly, LFRic is making some fundamental changes that impact the
underlying mesh and this links to I/O in terms of input/output file formats. Having
a concrete idea of what these file formats will be in the future enables LFRic to
inform future users, internal and external to the Met Office.

Although developing a bespoke I/O system was a possibility, the decision was
made early on to adopt an existing parallel I/O framework and leverage knowledge and
experience from the community; developing Earth System Models is becoming increasingly
complex and challenging for any one organisation to develop and maintain all the required
components~\cite{gmd-2017-186}.
As of early 2016, when the first scoping work for LFRic parallel I/O was being done,
there were several existing parallel I/O frameworks and XIOS was selected as the prime
candidate for evaluation~\cite{XIOSWiki}. A discussion of the other frameworks and why
they were not considered can be found in~\cite{Adams2018}.

XIOS was an obvious choice for LFRic as it was a mature framework, having had many years
of development and was already in use in the weather and climate domain; for example,
the Orchidee land surface model, the Dynamico dynamical core (part of the LMD-Z
Atmosphere model) and the NEMO ocean model. NEMO is used in Met Office coupled
models so there was some existing experience with XIOS. In terms of potential scalability,
XIOS had also been proved to run successfully in models running on the order of 10,000
cores. Crucially for climate models, XIOS works with the OASIS coupling framework that
is commonly used in coupled climate models including the UM. 

The key features of XIOS can be summarised as follows:
\begin{enumerate}
  \item Flexible definition of I/O workflows via external XML files;
  \item Client/Server architecture allowing asynchronous I/O servers on dedicated cores;
  \item Sophisticated ``in situ'' post-processing via workflows defined in the XML file(s). 
\end{enumerate}

XIOS is written in C++, but also provides a Fortran API to developers. It makes use of
MPI-IO and the NetCDF4 and HDF5 libraries and handles unstructured and regular grids.
XIOS is a client-server framework where I/O servers are asynchronous processes buffering
output requests from client processes. XIOS also has sophisticated post-processing
functionality - {\em e.g.} for regridding, and computing time series
and time averages. The output schedule and
file format are defined by an XML file which hides a lot of complexity from the user.

Prior to 2016, XIOS only supported read and write of NetCDF file formats that follow the CF
(Climate and Forecast) conventions~\cite{CFWeb}.  As previously explained in Section~\ref{sec:GH},
the LFRic implementation uses horizontally unstructured meshes and a FEM formulation,
where variables are held on different elements of the mesh (vertices, edges or faces).
Therefore, the I/O system needs to handle data structured in this way. The UGRID-NetCDF
format has been chosen as the main LFRic input mesh format and also output format for
diagnostics and model dumps~\cite{UgridSpec}. In the UGRID convention the topology of the
underlying unstructured mesh is stored as well as data, and data can be defined on any of
the mesh elements: vertices, edges, or faces. Working in collaboration with IPSL
(Institut Pierre Simon LaPlace), the ability to write the UGRID format was added to XIOS
in 2016 in order to support LFRic. 

XIOS has been integrated into LFRic via a lightweight I/O interface so
that (as far as possible) the underlying I/O framework is hidden and
could be replaced by an alternative.  The I/O interface is designed to
be compatible with the use of Fortran 2003 Object Orientation in
LFRic. The field object (see Figure~\ref{fig:objects}) contains read and write
interface declarations that use Fortran procedure pointers to define
I/O behaviours. When a field is created the behaviours are set by
pointing to a specific reading or writing method. Writing a field just
involves calling \verb+field%write_field()+ - {\em i.e.}, it is the
responsibility of the field to write itself and how the writing is
actually done is hidden.

More details of the experimental results of the LFRic-XIOS integration work can be
found in~\cite{Adams2018} where preliminary results on I/O scalability (up to 13824 
processors) and XIOS and Lustre performance tuning are presented. The conclusions of
this work were that XIOS is an effective parallel I/O framework that was technically
straightforward to integrate into the existing LFRic infrastructure via the XIOS
Fortran interface. Scaling experiments showed that XIOS performance was scalable with
respect to increasing numbers of cores, even with minimal tuning.

\section{\label{sec:psyclone}PSyclone}

As discussed earlier, PSyclone~\cite{psyclone} is a domain-specific
compiler which, given an algorithm and associated kernels, generates
the code for the middle, PSy layer. Currently, each of these layers
must be in Fortran and hence the compiler is embedded in
Fortran. This makes the generated code easy to understand for domain
scientists and means that existing debuggers and profilers can be used
as if it were any other hand written Fortran code. The choice of Fortran
is motivated in Section~\ref{sec:SoC}. However, there is nothing in the
approach to prevent other languages being targeted in the future. PSyclone
 itself is written in Python.

PSyclone is heavily influenced by the OP2 system~\citep{OP2,
  PYOP2}. However, PSyclone supports the specification of more than
one kernel in a parallel region of code, compared with a limit of one
for OP2, giving more scope for optimisation. Further, PSyclone takes
responsibility for distributed memory parallelisation whereas it must be
written manually by the application developer in OP2.

PSyclone is also designed to support a combination of kernels that are
1) written by the application developer and 2) provided by the
system. It also allows optimisations to be applied by an HPC expert
via a scripting interface, rather than necessarily automating the
process. By way of contrast, in the related
GridTools~\citep{grid_tools} and Firedrake~\citep{firedrake,fenics}
approaches, the application developer specifies the mathematical
operations (finite-difference stencils and finite element operations,
respectively) in a high-level language and optimised code is generated
automatically. Whilst this is a very powerful approach, it relies on
the high-level language capturing all possible operations and for the
system to produce highly optimised code automatically.

PSyclone, GridTools and Firedrake approaches are all based on the
concept of (various flavours of) a Domain-Specific Language (DSL) for
finite-difference and finite-element applications. This is distinct
from other, lower-level performance-portable abstractions, such as
Kokkos~\cite{KokkosGitHub} and OCCA~\citep{occa} where the aim is to
provide a language that permits an application developer to implement
a kernel once and have it compile to performant single-node code on a
range of multi- and many-core devices. These approaches compliment
DSL's, in that DSL's could make use of them to generate single-node
parallel code rather than using e.g. OpenMP or CUDA directly.

Lastly, CLAW~\citep{claw} is a Fortran source-to-source translation
tool designed to produce performance-portable Physics single-node
code. As such it is complimentary to PSyclone and could be used to
optimise PSyclone physics kernels that have been written by
application developers.

\begin{figure}
\centering\includegraphics[width=0.8\linewidth]{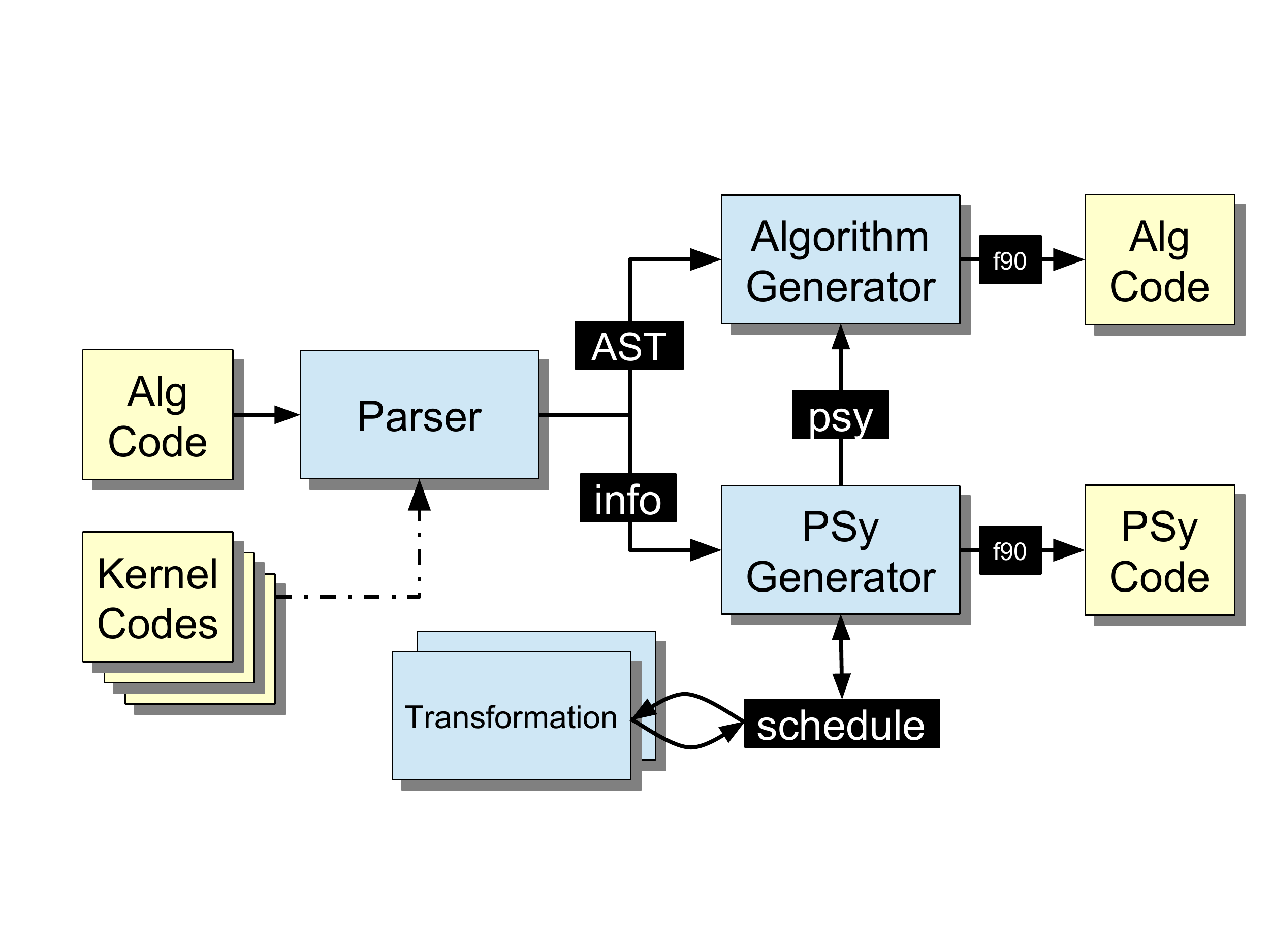}
\caption{\label{fig:psyclone-arch}Data flow within the PSyclone architecture.}
\end{figure}

Figure~\ref{fig:psyclone-arch} shows the data flow within the PSyclone
architecture. Starting from the left, the Algorithm is parsed (using
fparser, a pure Python implementation of a Fortran parser) and any
invoke procedure calls identified. The list of kernels (and their arguments)
associated with each of these calls is then stored. For each invoke,
the Fortran modules containing the kernel code (as identified by
the Fortran use statements) are parsed and the meta-data for each kernel
extracted.

PSyclone must then perform two tasks. Firstly, generate an internal
representation, known as a {\em schedule}, of a PSy layer routine for each invoke in
the Algorithm. If distributed memory is specified then this schedule
must include appropriate halo exchanges and global sums to ensure
correct execution. Secondly, PSyclone must modify the supplied Algorithm
code and replace each `call invoke' with a call to the corresponding
routine in the PSy layer. This latter task is achieved by modifying
the fparser-produced Abstract Syntax Tree (AST) of the Algorithm and
then re-generating Fortran from the new AST.

If no transformations are being applied (see below) to the PSy layer
schedule(s) then all that remains is for the PSy generator to create
Fortran code.  The steps involved in generating a vanilla
(un-optimised) PSy layer subroutine are as follows:
\begin{enumerate}
\item Generate the unique list of arguments required by all kernels
  in the invoke;
\item Generate Fortran which queries the LFRic infrastructure to get
  information on the sizes of the various iteration spaces (number of cells,
  vertical levels, dofs {\em etc.});
\item Generate Fortran which queries the LFRic infrastructure to get the
  various look-up tables ({\em e.g.} for identifying the dofs belonging to a
  given cell);
\item Generate Fortran which dereferences the various field and operator
  objects to get the necessary data arrays;
\item For each kernel in the invoke
  \begin{enumerate}
    \item Generate a loop over the correct iteration space (cells or dofs),
    \item Generate a call to the kernel subroutine with all necessary arguments.
  \end{enumerate}
\end{enumerate}

Since the algorithm deals with global objects
(fields, operators {\em etc.})  and a kernel deals with {\em e.g.} individual
columns of cells (see {\em e.g.} Section~\ref{sec:intro}), 
the PSy layer must also account for the fact that
the global objects are in fact decomposed over MPI processes by the
LFRic infrastructure. PSyclone must therefore ensure that global sums
are placed appropriately and that the halos of any fields read by a
given kernel are up-to-date before that kernel is executed. Initially
clean halos were implemented simply by generating code that, at
run-time, marked any field written to by a kernel as having a dirty
halo (using the LFRic infrastructure). Prior to each kernel call code
was also inserted that, for each field with a halo access, performed a
halo swap if the halo was dirty. Subsequently, dependence analysis has
been added and now only those halo exchanges that are known to be
required are inserted ~\cite{psyclone}.

\subsection{Transformations}

The large variety in existing computer architectures, the number of
different compilers and the fact that all of these are constantly
evolving means that it is simply not possible to write or generate a
single source code that will be (and continue to be) performance
portable~\cite{shallow_psykal, nemolite2d_psykal}.  It is also very
difficult to create a system that is automatically able to create
performant code for such a range of conditions. PSyclone therefore
seeks to be a tool for the HPC expert, enabling them to apply the
optimisations that, thanks to their experience and knowledge, they
know to be beneficial for a particular architecture and/or compiler.

The application of optimisations is achieved by applying
PSyclone-provided transformations to the schedule of an invoke.  For
instance, in order to parallelise a given schedule using OpenMP,
 PSyclone provides transformations to create a parallel region and
introduce various forms of work-sharing construct. However,
work-sharing of loops over cells for kernels which update quantities
on horizontally continuous function spaces will result in race
conditions where multiple threads attempt to write to the same dof on
a shared mesh entity. It
is therefore necessary to control the iteration space in order to
prevent these race conditions and PSyclone provides a loop-colouring
transformation for this purpose.

Although PSyclone transformations may be applied within a Python
interactive session, they will typically be used during the
compilation phase of a potentially large application. Therefore,
PSyclone allows (Python) transformation scripts to be supplied on the
command line. For example, to add OpenMP to the PSy layer, three
transformations need to be applied. Firstly, the colouring
transformation is applied to all loops in the schedule which contain
fields on horizontally continuous spaces that are modified, Secondly,
all loops, except loops over colours, have two OpenMP transformations
applied. This script, shown in Listing~\ref{lst:omp.py} in the
Appendix~\ref{app1}, has 17 lines of executable Python and is all that is
required to apply OpenMP to the whole model. Note, PSyclone supports
the application of transformations to any subset of a schedule and to
any set of schedules so may be used for general whole-code
optimisations and/or specific optimisations to particular code
regions.

Applying the OpenMP transformations to the invoke call described in
Section~\ref{sec:SoC} results in the following generated PSy layer
code shown in Listing~\ref{lst:PSy_omp}. 
\begin{lstlisting}[language=Fortran, numbers=left,caption={Code 
fragment of the generated PSy layer with OpenMP transformations},label={lst:PSy_omp}]
  CALL v_proxy%vspace%get_colours(ncolour, ncp_colour, cmap)
  DO colour=1,ncolour
    !$omp parallel default(shared), private(cell)
    !$omp do schedule(static)
    DO cell=1,ncp_colour(colour)
      CALL matrix_vector_code(cmap(colour, cell), nlayers, &
        v_proxy%data, s_proxy%data, mm_proxy%ncell_3d, &
        mm_proxy%local_stencil, ndf_any_space_1_v, &
        undf_any_space_1_v, &
        map_any_space_1_v(:,cmap(colour, cell)), &
        ndf_any_space_2_s, undf_any_space_2_s, &
        map_any_space_2_s(:,cmap(colour, cell)))
    END DO
    !$omp end do
    !$omp end parallel
  END DO
\end{lstlisting}
The transformation has resulted in a call to the LFRic infrastructure
to obtain the colouring information, a loop over the number of
colours, then the OpenMP workshare directives, which parallelise the
loop over the cells and the colour.

\section{\label{sec:Solver}Linear solvers and preconditioners}

Iterative solvers for large sparse linear systems of equations are
required in several places in the model. For example, since mass
matrices in the finite element discretisation are not diagonal, they
need to be inverted with a small number of iterations of a Krylov
subspace method. More importantly, the semi-implicit time-stepping
approach requires the solution of a very large sparse system for all
prognostic unknowns in each step of the non-linear Picard
iteration (Section~\ref{sec:sub:timestepping}). 
Since the system is ill-conditioned, it needs to be
preconditioned efficiently. This is achieved by the (approximate)
reduction to an elliptic system for the pressure, which itself is
preconditioned with a tensor-product multigrid algorithm~\cite{Borm2001} 
(see Section~\ref{sec:sub:preconditioner}). To increase
efficiency, the pressure preconditioner can be wrapped in its own
iterative solver for the Helmholtz system. Note that in contrast to
the approach already employed in the ENDGame model~\cite{QJ:QJ2235},
an outer iteration over the full system is still required due to the
non-diagonal nature of the finite element mass matrices. Altogether
this results in a rather complex solver.

\subsection{\label{sec:sub:solvinf}Solver infrastructure}

\begin{figure}
  \begin{center}
    \includegraphics[width=0.45\linewidth]{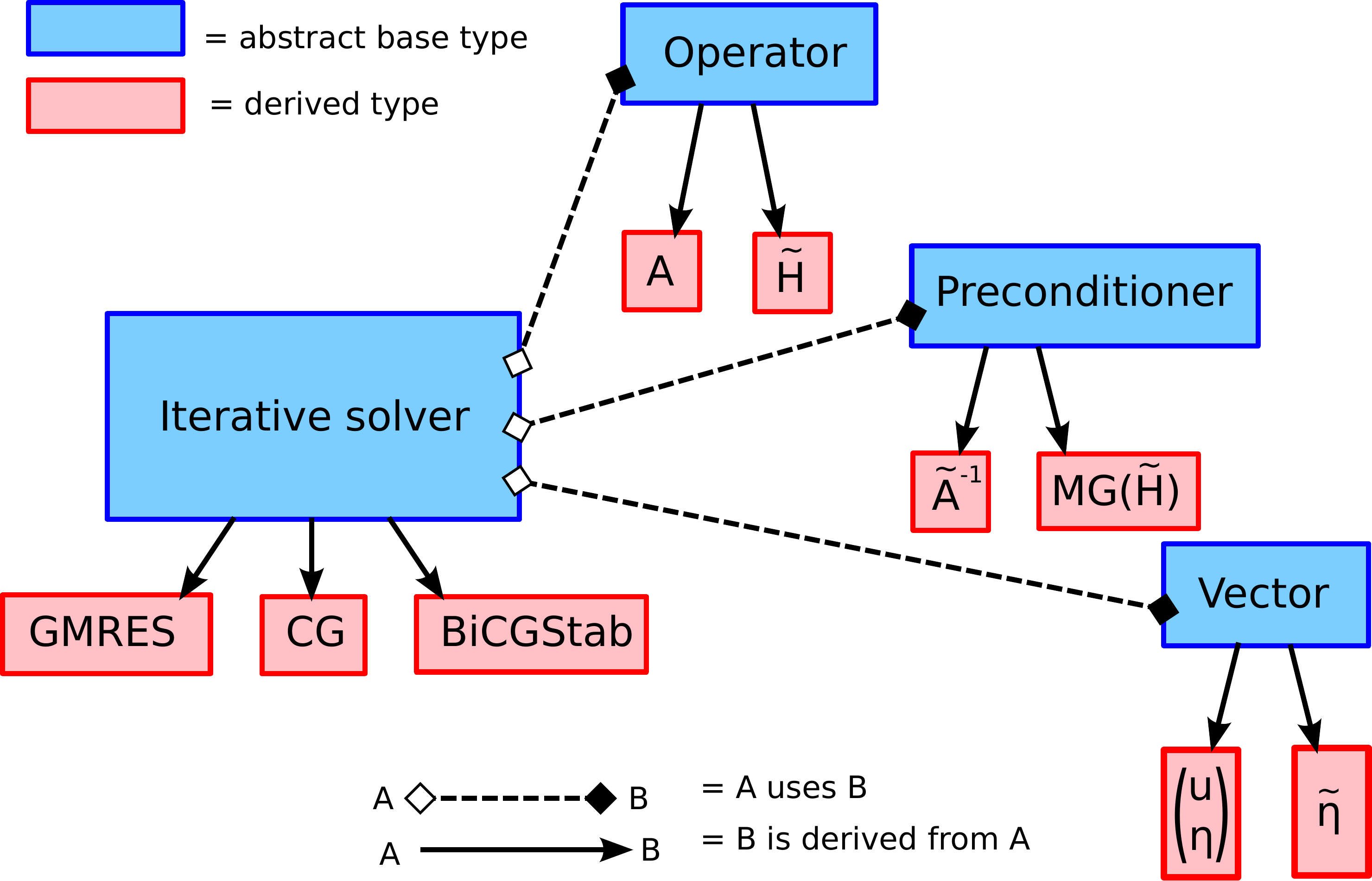}
    \hfill
    \includegraphics[width=0.45\linewidth]{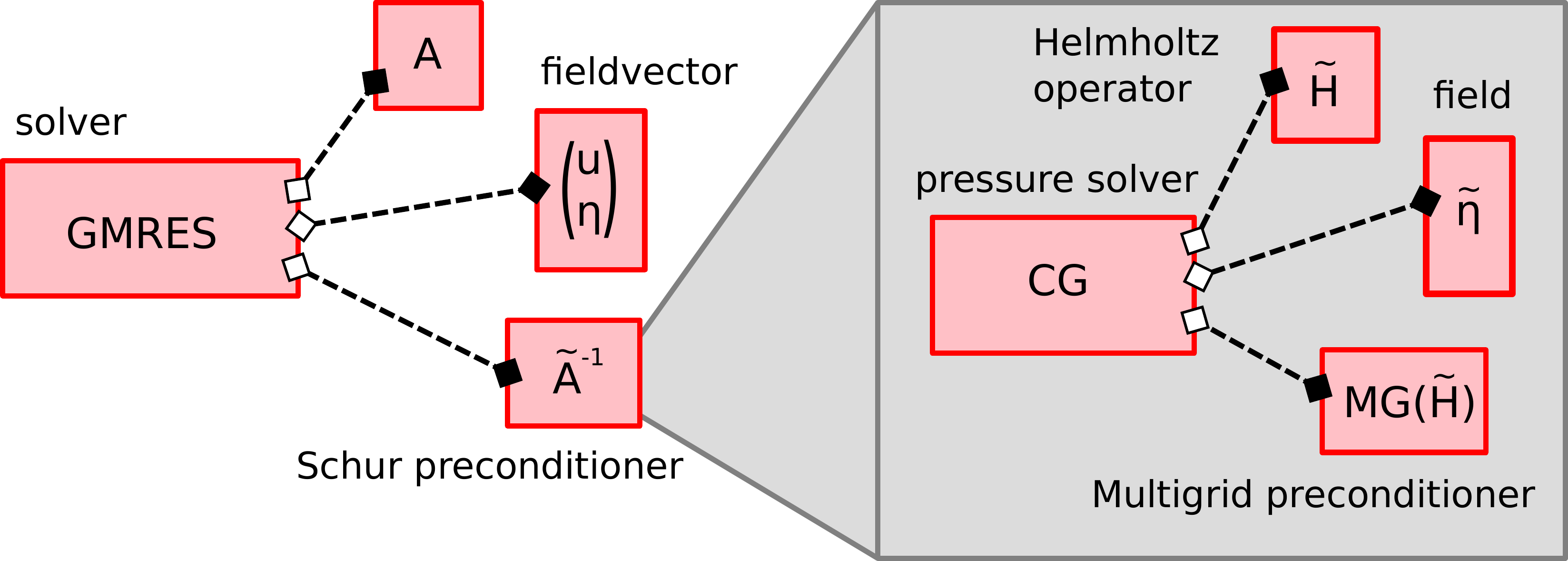}
    \caption{Derived class hierarchy for solvers and preconditioners
      in LFRic (left) and concrete implementation for the linear
      system in implicit time-stepping (right).}
    \label{fig:class_hierarchy}
  \end{center}
\end{figure}

To allow easy implementation of sophisticated nested iterative solvers
and preconditioners, a dedicated abstraction was developed by using
object-oriented features of Fortran 2003. This approach
follows similar design philosophies in widely used linear algebra
libraries such as PETSc~\cite{Balay1997,Balay2018} and DUNE-ISTL~\cite{Blatt2007}. 
More specifically, the implementation in LFRic uses
derived types which realise the following operations (see
Fig.~\ref{fig:class_hierarchy}, left):
\begin{itemize}
\item \textbf{Vector} types which support common linear algebra
  operations such as AXPY $y\mapsto y+\alpha x$ and dot products
  $x,y\mapsto s = \langle x,y\rangle$. The most important vector-type
  is \texttt{field\_vector}, which contains a collection of model
  fields;
\item \textbf{Linear operator} types which implement the operation $x\mapsto y=Ax$ for vectors $x$ and $y$;
\item \textbf{Preconditioners} which approximately solve the system $Px=b$ for a
  given right hand side $b$ and some operator $P\approx A$;
\item \textbf{Iterative solvers} which solve the system $Ax=b$ with a
  Krylov-subspace method for a given right hand side $b$. Currently
  supported solvers include Conjugate Gradient, GMRES and BiCGStab.
\end{itemize}
Each of those types is derived from an abstract base type. The
iterative solver types operate on generic vector types and are passed
preconditioner and linear operator objects which adhere to the
interface of their abstract base types.  This implies that only one
instance of a particular Krylov method has to be implemented in the
code. Apart from avoiding code duplication, this increases reliability
and maintainability, since only one instance of each solver has to be
developed and tested. In addition, it allows easy ``plug-and-play'' to
explore the space of all possible solver/preconditioner combinations
to achieve optimal performance.

To solve a particular problem, the user has to develop bespoke derived
types for the corresponding linear operator and preconditioners. Note
that the \texttt{apply()} methods of those derived types contain
invoke calls to kernels, which guarantees optimal performance of the entire
model. For example, to construct a solver for the implicit linear
system which is inverted in every semi-implicit time-step, the user
implements the following objects (see Fig.~\ref{fig:class_hierarchy},
right):
\begin{itemize}
\item A linear operator type which applies the full linear system to
  all components of a \texttt{field\_vector};
\item A preconditioner type which reduces the full linear system to
  the (approximate) Schur-complement in pressure space by lumping the
  velocity mass matrix; this preconditioner then calls the solver for
  the pressure (Helmholtz) system;
\item A linear operator type which applies the Helmholtz operator for
  the pressure system;
\item A preconditioner type which approximately inverts the Helmholtz
  operator (see Section~\ref{sec:sub:preconditioner}).
\end{itemize}
Once implemented, those linear operators/preconditioners need to be
passed to suitable existing linear solvers.

In addition to this more traditional approximate Schur-complement
approach for solving the full linear system, the development of
solvers based on a hybridisation approach is currently being 
explored~\cite{cockburnandjay2004,DBLP:journals/corr/abs-1802-00303}. 
Since an exact Schur-complement can be formed in this case, hybridisation
avoids the expensive iteration over the full linear system, and
potentially leads to significant improvements in performance.

\subsection{\label{sec:sub:preconditioner}Preconditioner}

To precondition the strongly anisotropic Helmholtz operator for the
pressure system, the tensor-product multigrid approach in~\cite{Borm2001} 
is being developed for use in LFRic. This is a more advanced solver than
the current tridiagonal vertical-only preconditioner in the ENDGame
model. The multigrid algorithm currently being developed in LFRic has been tested
extensively for representative elliptic equations in atmospheric
modelling in~\cite{Mueller2014,Dedner2016}, including a mixed-finite
element discretisation of a linear gravity wave system in~\cite{Mitchell2016}. 
The key idea to address the strong vertical
anisotropy due to the high-aspect ratio of the domain is to combine
vertical-only smoothing (line relaxation) with a horizontal multigrid
hierarchy. To allow the easy construction of the vertical-only
operators in the Schur-complement from the finite element
discretisation of the full equations, a suitable operator algebra was
developed in~\cite{Mitchell2016} and tested in the Firedrake
library. For horizontally discontinuous function spaces (such as the
pressure space, $Q_0^D$, and the vertical-only components of the $RT_0$ 
velocity space, see Section~\ref{sec:sub:spatial} for more details), 
operators can be partially assembled into a matrix type which 
stores all couplings within one vertical column. Matrices of this type can be
multiplied, added and, most importantly, inverted in the tridiagonal
solvers which realises the vertical line relaxation. This allows the
high-level construction of the building blocks required for the
Helmholtz solver and preconditioner. The same data structures were
implemented as derived Fortran 2003 types in the LFRic code and form
the building blocks of the tensor-product multigrid preconditioner for
the elliptic Helmholtz system.

Compared to simpler preconditioners, which do not combine vertical
line relaxation with a horizontal multigrid hierarchy, the
tensor-product multigrid approach typically reduces the solution time
of the pressure system by a factor of at least two
~\cite{Mueller2014,Mitchell2016}. However, since the Helmholtz system
contains a zero-order term, only a small number ($\approx 3-4$) of
multigrid levels is required (independent of the grid
resolution). This greatly increases scalability since it avoids global
couplings which arise on the coarsest level.

\section{\label{sec:scal}Scaling}

A key goal of 
LFRic and GungHo is scalability to meet run-time requirements. The use of an unstructured mesh from
GungHo and the solver construction described in
Section~\ref{sec:Solver} should enable the LFRic model to scale to a
very large degree of parallelism. However, compiling some of the
Fortran 2003 object-oriented constructs necessary to build such a solver
infrastructure is a challenging task for most compilers which has
delayed implementing such a solver. This remains a work in progress.

The February 2018 release of LFRic contains a simpler
solver. This employs a Krylov iterative GCR algorithm on the mixed
system of velocity, potential temperature, density and Exner
pressure. This is preconditioned by an approximate Schur-complement
for the Exner pressure which uses lumped mass matrices. The pressure
(Helmholtz) system is solved with BiCGstab and preconditioned by a
vertical only tridiagonal solver. Both mixed and Helmholtz solvers use
a relative tolerance of $10^{-6}$.

\begin{figure}
  \begin{center}
    \includegraphics[width=0.85\linewidth]{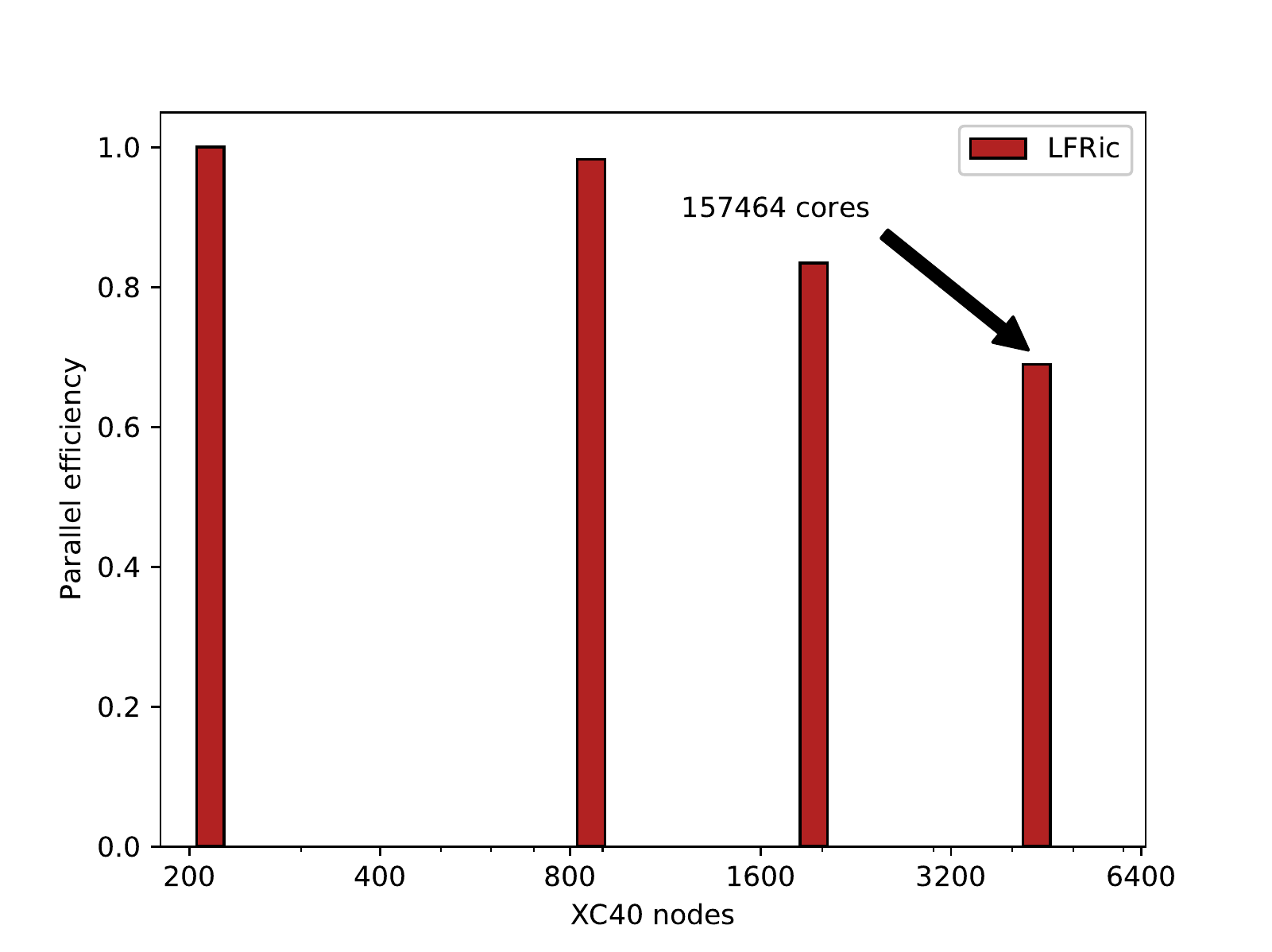}
    \caption{\label{fig:scale_PE}Strong scaling of the parallel efficiency of the LFRic 
      time-steps compared to 216 nodes. {\em N.B.} The x-axis shows a
      logarithmic scale}
  \end{center}
\end{figure}

Shown in Figure~\ref{fig:scale_PE} is 
the parallel efficiency of the LFRic time-step for a very high 
resolution model run. The mesh is a $C1944$ cubed-sphere with 30 
levels, where each panel has $1944 \times 1944$ cells. This roughly corresponds 
to a 5 km global resolution. The science configuration is the baroclinic wave 
test~\cite{qj.2241} with a 75 second time-step. The code was compiled 
to production level (\verb+-O3+) with the Intel 17 compiler. The model 
was run in hybrid mode with both MPI and OpenMP on the Met Office Cray 
XC40. Each node comprises of dual socket Intel Xeon (Broadwell) 
18-core processors. The affinity was set to be three MPI ranks per 
socket with six OpenMP threads per MPI rank. 

The model shows good strong scaling out to a very large number of nodes,
$4374$, which is $157366$ cores. Scaling starts to drop off at this
scale to just below $70\%$ compared to $216$ nodes. Here, the local
volume has been reduced to $12\times 12$ with only $30$ levels that is
only $4320$ dofs per unit of parallelism for the pressure space. The
ratio of communication to computation cost will become worse if the
problem is scaled further. The number of iterations for the
solver is relatively large and with multiple solves per time-step and
five global sums per iteration of the BiCGStab algorithm, the
cost of the global sums becomes prohibitive.

At this stage it would be difficult to make a meaningful comparison to
the performance of the current UM. The physical parameterisation
schemes are, as yet, mostly absent from LFRic. Moreover, the UM has
been highly optimised for the CPU architecture. The unstructured mesh
will give LFRic the algorithmic scaling advantage but the absolute
performance, {\em i.e.} wall-clock time, will still be
inferior. However, the new solver implementation will enable better
scaling by reducing the number of Krylov solver iterations required by
using a much more efficient preconditioner. This will also improve the
absolute performance, {\em i.e.} reduce run-time. Until the solver is
closer to being algorithmically optimal it does not make sense to
apply further computational optimisations.

\begin{figure}
  \begin{center}
\includegraphics[width=0.75\linewidth]{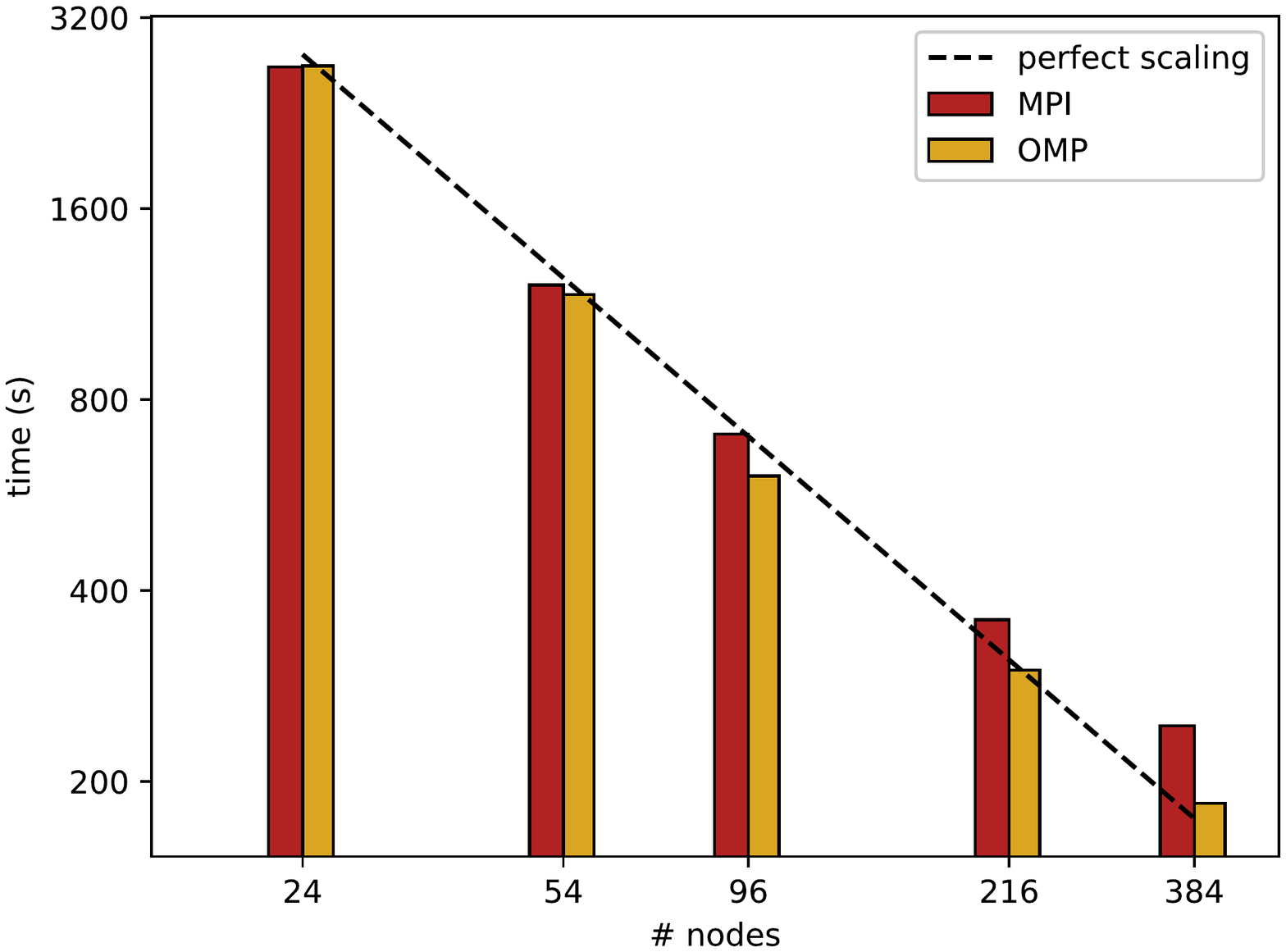}
    \caption{\label{fig:OMP_scale}a) Log-log plot of strong scaling of
      50 time-steps of the baroclinic wave problem on a cubed sphere
      mesh, C576, with 30 levels from the May 2018
      release. The red bars, labelled MPI, show MPI only, {\em i.e.}
      36 MPI ranks per node. The yellow bars, labelled OMP, show
      mixed-mode, with 6 MPI ranks per node and 6 OpenMP threads per
      MPI rank. The dashed line shows perfect scaling and is drawn to
      guide the eye.} 
  \end{center}
\end{figure}

Shown in Figure~\ref{fig:OMP_scale} is the effect of using OpenMP on 
scaling for 50 time-steps of the baroclinic wave problem, in this case
with the May 2018 release of LFRic. The code was
again compiled with the Intel 17 compiler. The model employs redundant
computation into the halos. Each processor computes the contribution
to dofs that reside on a shared mesh
entity (such as a face) which belong to a cell adjacent to a halo cell
from both the ``owned'' cell and the halo cell. Thus avoiding
communication that would otherwise be required to correctly calculate
the contribution of neighbour cells to that dof. Comparing MPI only to
the hybrid mode, each processor in the former regime has more work to
do in the case of redundant computation from the scaling of volume to
surface area. The mixed-mode has fewer cells to redundantly compute than
MPI only. Choosing redundant computation as a communication reduction
algorithm favours threading; hence OpenMP is faster and scales better.

\section{\label{sec:con}Discussion and Conclusion}

This paper describes the novel features of the design of the new LFRic
model and infrastructure, and the GungHo dynamical core within it. The
challenges of GungHo's higher order FEM approach on a horizontally
unstructured mesh essentially resulted in a requirement to develop a
new infrastructure from scratch. This requirement provided an
opportunity to develop a Domain Specific Language by separating the
concerns of the science code from the parallel code.  A domain
specific compiler called PSyclone has been developed to automatically
generate the OpenMP and MPI parallel code required to deploy the
application on an HPC machine.

DSLs also feature heavily in other work seeking to make weather and climate
models exascale ready. In contrast to this work, they are only part of the solution. For example,
MeteoSwiss have ported the COSMO model to run on GPUs, by
rewriting the dynamical core code using a DSL (STELLA). Other parts
of the code ({\em e.g.} physics) have been instrumented with compiler
directives~\cite{JSFI17}. CSCS (the Swiss National  Supercomputing
Centre) have made significant progress in porting the ICON dynamical
core to run on GPUs by instrumentation with OpenACC directives and
several physics schemes have been ported using a combination of
OpenACC and the CLAW source to source translation tool. To further
this effort, the ENIAC (ENable Icon on a heterogeneous ArChitecture)
project seeks to make the ICON model exascale ready using a similar
strategy of tools (the GridTools DSL and the CLAW source to source
translator) and OpenACC directives~\cite{ENIACDocs}.
The E3SM (Energy Exascale Earth System Model) project (a collaboration
between 7 National US laboratories)~\cite{E3SMDocs} has ported
HOMME (the dynamics/transport component of the Community Earth System
Model) to a single source version (HOMMEXX) able to run on MIC,
CPU and GPU architectures using Kokkos~\cite{KokkosGitHub}.

In summary, the design of LFRic has so far proven to be successful. The
introduction of PSyclone has enabled seamless switching from serial to
parallel running. Having initially being run in serial for many
months while the first version of PSyclone was developed, it was possible
to run in parallel (both MPI and OpenMP) on 220,000 cores of the Met Office
Cray XC40 within two weeks of the full PSyclone functionality
becoming available.

The GungHo dynamical core is routinely run in a range of science
configurations and several scientists are currently developing code within
the system. For the most part, scientists develop algorithms and kernels
within the existing infrastructure and with auto-generated PSy layer
code. For some newer developments the PSy layer code is hand-written
while new features are added to PSyclone. A clean separation between
the science code and the PSy layer code has been maintained
throughout. 

The need for PSyclone and the LFRic infrastructure to continuously
evolve in the light of new and changing science requirements
underlines an important point: the `separation of concerns' that has
enabled the science code to be kept isolated from the parallel systems
code does not mean that scientists and engineers can work in
isolation. In fact, the close ties between the scientists and software
engineers has been a key part of the success of the approach.

The scaling performance of the GungHo dynamical core is promising. While
significant improvements in the algorithmic design are required to
deliver better throughput of the model, the application (with I/O
capability switched off) scales well, with scaling tailing off at
around 160,000 cores due to the local volume of current configurations
becoming very small. One of the first improvements to the algorithmic
performance is anticipated to come from application of a geometric
multigrid solver within the new solver infrastructure discussed in the paper.

Currently, the generation of MPI and OpenMP code is supported by
PSyclone.  The approach of computing results into the halo redundantly
so as to reduce interprocessor communications helps the performance of
deployments with more OpenMP threads per MPI rank (for a given total
core count) because the larger domain of the high-thread runs means
that the relative size of the halos, and therefore the relative amount
of redundant computation, is smaller.  Moreover, PSyclone has been
developed with the ability to add transformations that increase the
amount of redundant computation so as to reduce the number of halo
swaps, to re-order kernels and to loop-fuse kernels to provide a
performance benefit. Work is also underway to support OpenACC
transformations which can then be applied without having to re-write
the science code.

The overall design has sought to carefully modularise functionality. A
benefit of this is to enable trialling of different externally sourced
libraries. An example of this was when the ESMF library was replaced by the
YAXT library in a short period of time.

Development of the core LFRic infrastructure and PSyclone is still in
its early days. LFRic is currently written to support any order of finite
element method on a range of different horizontally unstructured meshes.
Should the GungHo science converge on a particular mesh or set of meshes,
and on a particular FEM order it will likely be possible to refactor the
underlying LFRic data model to support more efficient access to the
data.

As well as continuing the technical capabilities of LFRic, currently,
sub-grid physics processes are being added to the GungHo dynamical
core implementation of LFRic with a view to steadily building up to a
fully-specified NWP model around 2020. Assuming the scientific and
computational performance results are acceptable it will likely go
head-to-head with the Met Office UM for 3-4 years while it is further
enhanced and optimised, eventually replacing the Unified Model
atmosphere at the core of the Met Office operations.

\section*{Acknowledgements}
The original designs for the computational infrastructure and the GungHo
formulation arose from the GungHo project. The contribution of GungHo
Consortium members as well as UM Partnership collaborators and GHASP team
members is gratefully acknowledged.

\section*{\label{app1}Appendix}

\begin{lstlisting}[language=python,caption={Python script for applying
transformations to a PSyclone schedule.},label={lst:omp.py}]
def trans(psy):
   ctrans = Dynamo0p3ColourTrans()
   otrans = Dynamo0p3OMPLoopTrans()
   oregtrans = OMPParallelTrans()
   # Loop over all of the invokes in the PSy object
   for invoke in psy.invokes.invoke_list:
       schedule = invoke.schedule
       # Colour loops over cells unless on discontinuous spaces
       for loop in schedule.loops():
           if loop.iteration_space == "cells" \
               and loop.field_space.orig_name \
                   not in DISCONTINUOUS_FUNCTION_SPACES:
               schedule, _ = ctrans.apply(loop)
       # Add OpenMP to loops unless they are over colours
       for loop in schedule.loops():
           if loop.loop_type != "colours":
               schedule, _ = oregtrans.apply(loop)
               schedule, _ = otrans.apply(loop, reprod=True)
   return psy
\end{lstlisting}

\newpage

%\bibliography{mibib.bib}

\begin{thebibliography}{10}
\expandafter\ifx\csname url\endcsname\relax
  \def\url#1{\texttt{#1}}\fi
\expandafter\ifx\csname urlprefix\endcsname\relax\def\urlprefix{URL }\fi
\expandafter\ifx\csname href\endcsname\relax
  \def\href#1#2{#2} \def\path#1{#1}\fi

\bibitem{gmd-10-1487-2017}
D.~Walters, I.~Boutle, M.~Brooks, T.~Melvin, R.~Stratton, S.~Vosper, H.~Wells,
  K.~Williams, N.~Wood, T.~Allen, A.~Bushell, D.~Copsey, P.~Earnshaw,
  J.~Edwards, M.~Gross, S.~Hardiman, C.~Harris, J.~Heming, N.~Klingaman,
  R.~Levine, J.~Manners, G.~Martin, S.~Milton, M.~Mittermaier, C.~Morcrette,
  T.~Riddick, M.~Roberts, C.~Sanchez, P.~Selwood, A.~Stirling, C.~Smith,
  D.~Suri, W.~Tennant, P.~L. Vidale, J.~Wilkinson, M.~Willett, S.~Woolnough,
  P.~Xavier, \href{https://www.geosci-model-dev.net/10/1487/2017/}{{The Met
  Office Unified Model Global Atmosphere 6.0/6.1 and JULES Global Land 6.0/6.1
  configurations}}, Geoscientific Model Development 10~(4) (2017) 1487--1520.
\newblock \href {http://dx.doi.org/10.5194/gmd-10-1487-2017}
  {\path{doi:10.5194/gmd-10-1487-2017}}.
\newline\urlprefix\url{https://www.geosci-model-dev.net/10/1487/2017/}

\bibitem{gmd-2017-186}
B.~N. Lawrence, M.~Rezny, R.~Budich, P.~Bauer, J.~Behrens, M.~Carter,
  W.~Deconinck, R.~Ford, C.~Maynard, S.~Mullerworth, C.~Osuna, A.~Porter,
  K.~Serradell, S.~Valcke, N.~Wedi, S.~Wilson,
  \href{https://www.geosci-model-dev.net/11/1799/2018/}{Crossing the {Chasm:
  How} to develop weather and climate models for next generation computers?},
  Geosci. Model Dev. 11 (2018) 1799--1821.
\newblock \href {http://dx.doi.org/10.5194/gmd-11-1799-2018}
  {\path{doi:10.5194/gmd-11-1799-2018}}.
\newline\urlprefix\url{https://www.geosci-model-dev.net/11/1799/2018/}

\bibitem{GHP1_CSR}
R.~Ford, M.~Glover, D.~Ham, C.~Maynard, S.~Pickles, G.~Riley,
  \href{https://www.metoffice.gov.uk/binaries/content/assets/mohippo/pdf/8/o/frtr587tagged.pdf}{{GungHo
  Phase 1: Computational} science recommendations}, Technical report, Met
  Office (2013).
\newline\urlprefix\url{https://www.metoffice.gov.uk/binaries/content/assets/mohippo/pdf/8/o/frtr587tagged.pdf}

\bibitem{melvin2018}
T.~Melvin, T.~Benacchio, B.~Shipway, N.~Wood, J.~Thuburn, C.~Cotter, A mixed
  finite-element, semi-implicit finite-volume discretisation for atmospheric
  dynamics: {C}artesian geometry, Q. J. Roy. Meteorol. Soc.In preparation.

\bibitem{Lynch2006}
P.~Lynch, The Emergence of Numerical Weather Prediction: {Richardson's} Dream,
  Cambridge University Press, 2006.

\bibitem{ForecastFactory}
P.~Lynch, Richardson's fantastic forecast factory, european meteorological
  society website, \url{https://www.emetsoc.org/resources/rff/}, accessed 12
  October 2018.

\bibitem{NairEtal2005}
R.~D. Nair, S.~J. Thomas, R.~D. Loft,
  \href{https://doi.org/10.1175/MWR2890.1}{A discontinuous {Galerkin} transport
  scheme on the cubed sphere}, Monthly Weather Review 133~(4) (2005) 814--828.
\newblock \href {http://dx.doi.org/10.1175/MWR2890.1}
  {\path{doi:10.1175/MWR2890.1}}.
\newline\urlprefix\url{https://doi.org/10.1175/MWR2890.1}

\bibitem{staniforth2012}
A.~Staniforth, J.~Thuburn, Horizontal grids for global weather prediction and
  climate models: {A} review, Q. J. Roy. Meteorol. Soc. 138 (2012) 1--26.
\newblock \href {http://dx.doi.org/10.1002/qj.958} {\path{doi:10.1002/qj.958}}.

\bibitem{MacDonaldEtal2011}
A.~E. MacDonald, J.~Middlecoff, T.~Henderson, J.-L. Lee,
  \href{https://doi.org/10.1177/1094342010385019}{A general method for modeling
  on irregular grids}, The International Journal of High Performance Computing
  Applications 25~(4) (2011) 392--403.
\newblock \href {http://dx.doi.org/10.1177/1094342010385019}
  {\path{doi:10.1177/1094342010385019}}.
\newline\urlprefix\url{https://doi.org/10.1177/1094342010385019}

\bibitem{cotter2012}
C.~J. Cotter, J.~Shipton, Mixed finite elements for numerical weather
  prediction, J. Comput. Phys. 231 (2012) 7076--7091.

\bibitem{natale2016}
A.~Natale, J.~Shipton, C.~J. Cotter, Compatible finite element spaces for
  geophysical fluid dynamics, Dynam. Stat. Climate Sys. 1~(1).
\newblock \href {http://dx.doi.org/10.1093/climsys/dzw005}
  {\path{doi:10.1093/climsys/dzw005}}.

\bibitem{charney1953numerical}
J.~G. Charney, N.~Phillips, Numerical integration of the quasi-geostrophic
  equations for barotropic and simple baroclinic flows, J. Meteor. 10~(2)
  (1953) 71--99.

\bibitem{arakawa1977computational}
A.~Arakawa, V.~R. Lamb,
  \href{http://www.sciencedirect.com/science/article/pii/B9780124608177500094}{Computational
  design of the basic dynamical processes of the {UCLA} general circulation
  model}, in: J.~Chang (Ed.), General Circulation Models of the Atmosphere,
  Vol.~17 of Methods in Computational Physics: Advances in Research and
  Applications, Elsevier, 1977, pp. 173--265.
\newblock \href
  {http://dx.doi.org/https://doi.org/10.1016/B978-0-12-460817-7.50009-4}
  {\path{doi:https://doi.org/10.1016/B978-0-12-460817-7.50009-4}}.
\newline\urlprefix\url{http://www.sciencedirect.com/science/article/pii/B9780124608177500094}

\bibitem{boffi2013}
D.~Boffi, F.~Brezzi, M.~Fortin, Mixed Finite Element Methods and Applications,
  Vol.~44, Springer, 2013.

\bibitem{Leonard1996}
B.~P. Leonard, A.~P. Lock, M.~K. MacVean, Conservative explicit
  unrestricted-time-step multidimensional constancy-preserving advection
  schemes, Mon. Weather Rev. 124~(11) (1996) 2588--2606.
\newblock \href
  {http://dx.doi.org/10.1175/1520-0493(1996)124{<}2588:CEUTSM{>}2.0.CO;2}
  {\path{doi:10.1175/1520-0493(1996)124{<}2588:CEUTSM{>}2.0.CO;2}}.

\bibitem{QJ:QJ2235}
N.~Wood, A.~Staniforth, A.~White, T.~Allen, M.~Diamantakis, M.~Gross,
  T.~Melvin, C.~Smith, S.~Vosper, M.~Zerroukat, J.~Thuburn,
  \href{http://dx.doi.org/10.1002/qj.2235}{An inherently mass-conserving
  semi-implicit semi-{Lagrangian} discretization of the deep-atmosphere global
  non-hydrostatic equations}, Q. J. Roy. Meteorol. Soc. 140~(682) (2014)
  1505--1520.
\newblock \href {http://dx.doi.org/10.1002/qj.2235}
  {\path{doi:10.1002/qj.2235}}.
\newline\urlprefix\url{http://dx.doi.org/10.1002/qj.2235}

\bibitem{dennard}
R.~H. Dennard, F.~H. Gaensslen, V.~L. Rideout, E.~Bassous, A.~R. LeBlanc,
  Design of ion-implanted {MOSFET's} with very small physical dimensions, IEEE
  J. Solid-State Circuits 9~(5) (1974) 256--268.
\newblock \href {http://dx.doi.org/10.1109/JSSC.1974.1050511}
  {\path{doi:10.1109/JSSC.1974.1050511}}.

\bibitem{ESMFDocs}
{ESMF} project website, \url{http://www.earthsystemmodeling.org}, accessed 23
  July 2018.

\bibitem{YAXTDocs}
{YAXT} project website, \url{https://www.dkrz.de/redmine/projects/yaxt},
  accessed 20 July 2018.

\bibitem{DECONINCK2017188}
W.~Deconinck, P.~Bauer, M.~Diamantakis, M.~Hamrud, C.~K\"{u}hnlein, P.~Maciel,
  G.~Mengaldo, T.~Quintino, B.~Raoult, P.~K. Smolarkiewicz, N.~P. Wedi,
  \href{http://www.sciencedirect.com/science/article/pii/S0010465517302138}{{Atlas
  : A} library for numerical weather prediction and climate modelling}, Comput.
  Phy. Comm. 220 (2017) 188 -- 204.
\newblock \href {http://dx.doi.org/https://doi.org/10.1016/j.cpc.2017.07.006}
  {\path{doi:https://doi.org/10.1016/j.cpc.2017.07.006}}.
\newline\urlprefix\url{http://www.sciencedirect.com/science/article/pii/S0010465517302138}

\bibitem{XIOSWiki}
{XIOS} wiki, \url{http://forge.ipsl.jussieu.fr/ioserver}, accessed 16 July
  2018.

\bibitem{Adams2018}
S.~V. Adams, O.~Abramkina, Y.~Meurdesoif, M.~Rezny, Parallel {IO} in the
  {LFRic} infrastructure, in: S.~Bassini, M.~Danelutto, P.~Dazzi, G.~R.
  Joubert, F.~Peters (Eds.), Parallel Computing is Everywhere, IOS Press, 2018,
  pp. 485--494.

\bibitem{CFWeb}
{CF Conventions and Metadata}, \url{http://cfconventions.org/}, accessed 12
  October 2018.

\bibitem{UgridSpec}
{UGRID Convention V1.0},
  \url{http://ugrid-conventions.github.io/ugrid-conventions}, accessed 16 July
  2018.

\bibitem{psyclone}
R.~W. Ford, A.~R. Porter, {PSyclone:} a domain-specific compiler for
  finite-element/volume/difference earth-system modelsIn preparation.

\bibitem{OP2}
C.~Bertolli, A.~Betts, G.~Mudalige, M.~Giles, P.~Kelly, Design and performance
  of the op2 library for unstructured mesh applications, in: M.~e.~a. Alexander
  (Ed.), Euro-Par 2011: Parallel Processing Workshops, Vol. 7155 of Lecture
  Notes in Computer Science, Springer Berlin Heidelberg, 2012, pp. 191--200.

\bibitem{PYOP2}
F.~Rathgeber, G.~R. Markall, L.~Mitchell, N.~Loriant, D.~A. Ham, C.~Bertolli,
  P.~H.~J. Kelly, \href{http://dx.doi.org/10.1109/SC.Companion.2012.134}{Pyop2:
  A high-level framework for performance-portable simulations on unstructured
  meshes}, in: Proceedings of the 2012 SC Companion: High Performance
  Computing, Networking Storage and Analysis, SCC '12, IEEE Computer Society,
  Washington, DC, USA, 2012, pp. 1116--1123.
\newblock \href {http://dx.doi.org/10.1109/SC.Companion.2012.134}
  {\path{doi:10.1109/SC.Companion.2012.134}}.
\newline\urlprefix\url{http://dx.doi.org/10.1109/SC.Companion.2012.134}

\bibitem{grid_tools}
T.~Gysi, C.~Osuna, O.~Fuhrer, M.~Bianco, T.~Schulthess, in: Proceedings of the
  International Conference for High Performance Computing, Networking, Storage
  and Analysis, SC ’15, ACM, 2015, pp. 1--41.

\bibitem{firedrake}
F.~Rathgeber, D.~A. Ham, L.~Mitchell, M.~Lange, F.~Luporini, A.~T. McRae, G.-T.
  Bercea, G.~R. Markall, P.~H. Kelly,
  \href{http://arxiv.org/abs/1501.01809}{Firedrake: automating the finite
  element method by composing abstractions}, Submitted to ACM TOMS\href
  {http://arxiv.org/abs/1501.01809} {\path{arXiv:1501.01809}}.
\newline\urlprefix\url{http://arxiv.org/abs/1501.01809}

\bibitem{fenics}
A.~Logg, K.-A. Mardal, G.~N. Wells, et~al., Automated Solution of Differential
  Equations by the Finite Element Method, Springer, 2012.
\newblock \href {http://dx.doi.org/10.1007/978-3-642-23099-8}
  {\path{doi:10.1007/978-3-642-23099-8}}.

\bibitem{KokkosGitHub}
{Kokkos GitHub} repository, \url{https://github.com/kokkos/kokkos}, accessed 30
  July 2018.

\bibitem{occa}
D.~Medina, A.~St-Cyr, T.~Warburton, High-order finite-differences on
  multi-threaded architectures using occa., in: R.~Kirby, M.~Berzins,
  J.~Hesthaven (Eds.), Spectral and High Order Methods for Partial Differential
  Equations ICOSAHOM 2014., Vol. 106 of Lecture Notes in Computational Science
  and Engineering, Springer, 2015, pp. 365--373.

\bibitem{claw}
V.~Clement, S.~Ferrachat, O.~Fuhrer, X.~Lapillonne, C.~E. Osuna, R.~Pincus,
  J.~Rood, W.~Sawyer, \href{http://doi.acm.org/10.1145/3218176.3218226}{The
  claw dsl: Abstractions for performance portable weather and climate models},
  in: Proceedings of the Platform for Advanced Scientific Computing Conference,
  PASC '18, ACM, New York, NY, USA, 2018, pp. 2:1--2:10.
\newblock \href {http://dx.doi.org/10.1145/3218176.3218226}
  {\path{doi:10.1145/3218176.3218226}}.
\newline\urlprefix\url{http://doi.acm.org/10.1145/3218176.3218226}

\bibitem{shallow_psykal}
A.~R. Porter, R.~W. Ford, M.~Ashworth, G.~D. Riley, M.~Modani, Towards
  compiler-agnostic performance in finite-difference codes, in: G.~R. Joubert,
  H.~Leather, M.~Parsons, F.~Peters, M.~Sawyer (Eds.), Parallel Computing: On
  the Road to Exascale, Vol.~27, IOS Press, Amsterdam, New York, Tokyo, 2016,
  pp. 647--658.

\bibitem{nemolite2d_psykal}
A.~Porter, J.~Appleyard, M.~Ashworth, R.~Ford, J.~Holt, H.~Liu, G.~Riley,
  \href{https://www.geosci-model-dev-discuss.net/gmd-2017-150/}{Portable multi-
  and many-core performance for finite difference codes; {A}pplication to the
  free-surface component of {NEMO}}, Geosci. Model Dev. Discussions 2017 (2017)
  1--27.
\newblock \href {http://dx.doi.org/10.5194/gmd-2017-150}
  {\path{doi:10.5194/gmd-2017-150}}.
\newline\urlprefix\url{https://www.geosci-model-dev-discuss.net/gmd-2017-150/}

\bibitem{Borm2001}
S.~B{\"o}rm, R.~Hiptmair, Analysis of tensor product multigrid, Numer.
  Algorithms 26~(3) (2001) 219--234.

\bibitem{Balay1997}
S.~Balay, W.~D. Gropp, L.~C. McInnes, B.~F. Smith, Efficient management of
  parallelism in object oriented numerical software libraries, in: E.~Arge,
  A.~M. Bruaset, H.~P. Langtangen (Eds.), Modern Software Tools in Scientific
  Computing, Birkh{\"{a}}user Press, 1997, pp. 163--202.

\bibitem{Balay2018}
S.~Balay, S.~Abhyankar, M.~F. Adams, J.~B.~P. Brune, K.~Buschelman, L.~Dalcin,
  V.~Eijkhout, W.~D. Gropp, D.~Kaushik, M.~G. Knepley, D.~A. May, L.~C.
  McInnes, R.~T. Mills, T.~Munson, K.~Rupp, P.~Sanan, B.~F. Smith, S.~Zampini,
  H.~Zhang, H.~Zhang, \href{http://www.mcs.anl.gov/petsc}{{PETS}c {W}eb page},
  \url{http://www.mcs.anl.gov/petsc}, accessed 25 July 2018.
\newline\urlprefix\url{http://www.mcs.anl.gov/petsc}

\bibitem{Blatt2007}
M.~Blatt, P.~Bastian, The iterative solver template library, in:
  B.~K{\aa}gstr{\"{o}}m, E.~Elmroth, J.~Dongarra, J.~Wa{\'s}niewski (Eds.),
  Applied Parallel Computing. State of the Art in Scientific Computing,
  Springer, 2007, pp. 666--675.
\newblock \href {http://dx.doi.org/10.1007/978-3-540-75755-9{\_}82}
  {\path{doi:10.1007/978-3-540-75755-9{\_}82}}.

\bibitem{cockburnandjay2004}
B.~Cockburn, J.~Gopalakrishnan, A characterization of hybridized mixed methods
  for second order elliptic problems, SIAM Journal on Numerical Analysis 42~(1)
  (2004) 283--301.

\bibitem{DBLP:journals/corr/abs-1802-00303}
T.~H. Gibson, L.~Mitchell, D.~A. Ham, C.~J. Cotter,
  \href{http://arxiv.org/abs/1802.00303}{A domain-specific language for the
  hybridization and static condensation of finite element methods}, CoRR
  abs/1802.00303.
\newblock \href {http://arxiv.org/abs/1802.00303} {\path{arXiv:1802.00303}}.
\newline\urlprefix\url{http://arxiv.org/abs/1802.00303}

\bibitem{Mueller2014}
E.~H. M{\"u}ller, R.~Scheichl, Massively parallel solvers for elliptic partial
  differential equations in numerical weather and climate prediction, Q. J.
  Roy. Meteorol. Soc. 140~(685) (2014) 2608--2624.

\bibitem{Dedner2016}
A.~Dedner, E.~M{\"u}ller, R.~Scheichl, Efficient multigrid preconditioners for
  atmospheric flow simulations at high aspect ratio, Int. J. Numer. Meth. F.
  80~(1) (2016) 76--102.

\bibitem{Mitchell2016}
L.~Mitchell, E.~H. M{\"u}ller, High level implementation of geometric multigrid
  solvers for finite element problems: {Applications} in atmospheric modelling,
  J. Comput. Phys. 327 (2016) 1--18.

\bibitem{qj.2241}
P.~A. Ullrich, T.~Melvin, C.~Jablonowski, A.~Staniforth, A proposed baroclinic
  wave test case for deep- and shallow-atmosphere dynamical cores, Q. J. Roy.
  Meteorol. Soc. 140~(682) (2013) 1590--1602.
\newblock \href {http://dx.doi.org/10.1002/qj.2241}
  {\path{doi:10.1002/qj.2241}}.

\bibitem{JSFI17}
O.~Fuhrer, C.~Osuna, X.~Lapillonne, T.~Gysi, B.~Cumming, M.~Bianco, A.~Arteaga,
  T.~Schulthess, \href{http://superfri.org/superfri/article/view/17}{Towards a
  performance portable, architecture agnostic implementation strategy for
  weather and climate models}, Supercomp. Front. Innov. 1~(1).
\newblock \href {http://dx.doi.org/10.14529/jsfi140103}
  {\path{doi:10.14529/jsfi140103}}.
\newline\urlprefix\url{http://superfri.org/superfri/article/view/17}

\bibitem{ENIACDocs}
{ENIAC} project website,
  \url{https://www.pasc-ch.org/projects/2017-2020/eniac-enabling-the-icon-model-on-heterogeneous-architectures/},
  accessed 30 July 2018.

\bibitem{E3SMDocs}
{E3SM} project website,
  \url{https://climatemodeling.science.energy.gov/projects/energy-exascale-earth-system-model},
  accessed 30 July 2018.

\end{thebibliography}

\end{document}